\documentclass[aps,twocolumn,prx,amsfonts,showpacs,superscriptaddress,longbibliography]{revtex4-2}
\usepackage[utf8]{inputenc}
\usepackage{graphicx}
\usepackage{bm}
\usepackage{hyperref}
\usepackage{amsmath}
\usepackage{amssymb}
\usepackage[table]{xcolor}
\usepackage{array}
\usepackage{multirow}
\usepackage[caption=false]{subfig}
\usepackage{physics}
\usepackage{booktabs}
\usepackage{soul}
\usepackage{amsthm}
\usepackage{bbm}
\usepackage{multirow}
\usepackage[colorinlistoftodos]{todonotes}
\usepackage[scr=boondoxo,frak=boondox,scrscaled=1.05]{mathalfa}
\usepackage{amsmath}

\def\ket#1{|#1\rangle}

\def\r{{\boldsymbol{r}}}

\def\k{{\boldsymbol{k}}}
\def\p{{\boldsymbol{p}}}

\def\q{{\boldsymbol{q}}}

\def\G{{\boldsymbol{G}}}
\def\Q{{\boldsymbol{Q}}}

\def\a{{\boldsymbol{a}}}

\def\K{{\boldsymbol{K}}}

\usepackage{ulem}

\begin{document}

\title{Rhombohedral Graphene: A Tale of Many Crystals}

\author{Ahmed Abouelkomsan}
\thanks{These authors contributed equally to this work.}
\affiliation{Department of Physics, Massachusetts Institute of Technology, Cambridge, Massachusetts 02139, USA}

\author{Filippo Gaggioli}
\thanks{These authors contributed equally to this work.}
\affiliation{Department of Physics, Massachusetts Institute of Technology, Cambridge, Massachusetts 02139, USA}

\author{Daniele Guerci}
\thanks{These authors contributed equally to this work.}
\affiliation{Department of Physics, Massachusetts Institute of Technology, Cambridge, Massachusetts 02139, USA}

\author{Liang Fu}
\affiliation{Department of Physics, Massachusetts Institute of Technology, Cambridge, Massachusetts 02139, USA}
\begin{abstract}
Experiments on rhombohedral graphene have uncovered an extraordinary wealth of correlated quantum phases—from chiral superconductors to electronic crystals—all within a single family of atomically thin materials. %Understanding the organizing principles of this phase diagram is a central challenge. 
Here, we introduce a simple indicator, derived from the noninteracting band dispersion, that identifies strongly correlated regions in the phase diagram of rhombohedral graphene as a function of carrier density and displacement field. We develop a neural-network variational Monte Carlo method, combined with Hartree-Fock theory, to solve the interacting ground states. Our calculation reveals a variety of electron crystals with no classical analog. These include, at increasing density: Wigner crystal, self-doped Wiger crystal, as well as ``anticrystal''---a lattice of holes in an electron liquid. We discuss their experimental manifestations and possible connection to superconductivity.       
\end{abstract}
\maketitle

\textit{Introduction ---}  Rhombohedral graphene has emerged as a most versatile platform for strongly correlated quantum matter. 
Within a single family of atomically thin carbon multilayers, experiments have now uncovered an exceptional range of distinct quantum phases including integer and fractional topological phases~\cite{han2024correlated,Lu2024Feb,lu_extended_2025,xie2025tunablefractionalcherninsulators,Choi_2025,Aronson2025}, chiral superconductors~\cite{han2025signatures,dutta2026reconfigurablechiralsuperconductivity,nguyen2025hierarchysuperconductivitytopologicalcharge,qin2026stripeordermetallicsuperconducting,hua2026multiknobswitchablechiralsuperconductivity,nguyen2026coexistingchargedensitywave}, and insulating and metallic electronic crystals~\cite{han2026evidencemetallicwignercrystal,zhou2026competingordersdrivenwigner}. 
While the phase diagram tuned by carrier density and displacement field is remarkably rich,  this richness also poses a central challenge:  identifying the organizing principles that control the emergence, competition, and coexistence of these interaction-induced phases.

A key feature of rhombohedral graphene is that the displacement field reshapes its low-energy band structure and can flatten the bottom of the band, leading to strong correlation effects. 
In particular, some of these chiral superconducting states appear near insulating states at lower density~\cite{han2025signatures}, which are believed to be electron Wigner crystals. The complex phase diagram in rhombohedral graphene as a function of carrier density $n$ and displacement field $D$, as well as its evolution with layer thickness, calls for theoretical understanding.

%Understanding these crystalline phases, as well as the competition and possible interplay between them, is therefore essential for developing a unified picture of the correlated phase diagram.

% A common feature of rhombohedral multilayer graphene phase diagrams, recently pointed out by some of us~\cite{geier2024chiraltopologicalsuperconductivityisospin}, is the emergence of an electronic crystal within this strongly correlated regime, separating two superconducting domes, SC1 and SC2.
% Clarifying the nature of this intervening crystalline region is therefore essential for understanding the organization of the phase diagram in this family of materials.

In this work, we find a useful guide for navigating the phase diagram of rhombohedral graphene, and present a comprehensive study of its $(n, D)$ phase diagram using state-of-the-art neural-network variational Monte Carlo and Hartree-Fock theory. Specifically, we introduce a simple indicator of strongly correlated regions in the phase diagram based on the noninteracting energy dispersion.  Guided by this indicator, our large-scale numerical simulations find a plethora of crystalline states in different regions of the phase diagram, enabled by the tunable fermiology of rhombohedral graphene and its interplay with Coulomb interaction.

Besides the standard Wigner crystal, we predict the existence of  quantum crystalline phases without classical analog, in the regime of annular Fermi sea. %competition of interaction and kinetic energy effects.
These include i) %a nodal Wigner crystal, characterized by charge depletion along a ring centered at the triangular-lattice centers, ii) 
a  self-doped crystal with non-integer filling of the unit cell, %number of peaks that differs from the number of charges in the system 
and ii) an ``anticrystal'' %where the crystalline order is carried by 
featuring a triangular lattice of holes (rather than localized electrons). %, which emerge in proximity of a commensurate $2:1$ filling ratio between the electron- and hole-like Fermi surfaces. 

\par
\textit{Fermiology of Rhombohedral Graphene ---}  In rhombohedral multilayer graphene, near charge neutrality, the low-energy bands originate from sublattice polarized states localized on the top and bottom layers \cite{Koshino_2009, Zhang_MacDonald_2010}. 
% Importantly, at low energy, these bands realize a higher-order dispersion whose leading form is dependent on number of layers $L $, $\varepsilon(\k) \propto |\k|^L $. 
Importantly, these low-energy states are coupled through a higher-order tunneling process whose order is determined by the number ($L$) of interlayer hopping events required to connect them, resulting in a coupling that scales as $|\k|^L$. 
A perpendicular displacement field $D$ induces a potential-energy imbalance between these layer-polarized states, opening a gap and providing a direct knob to tune the band dispersion. 

\begin{figure}
    \centering
    \includegraphics[width=\linewidth]{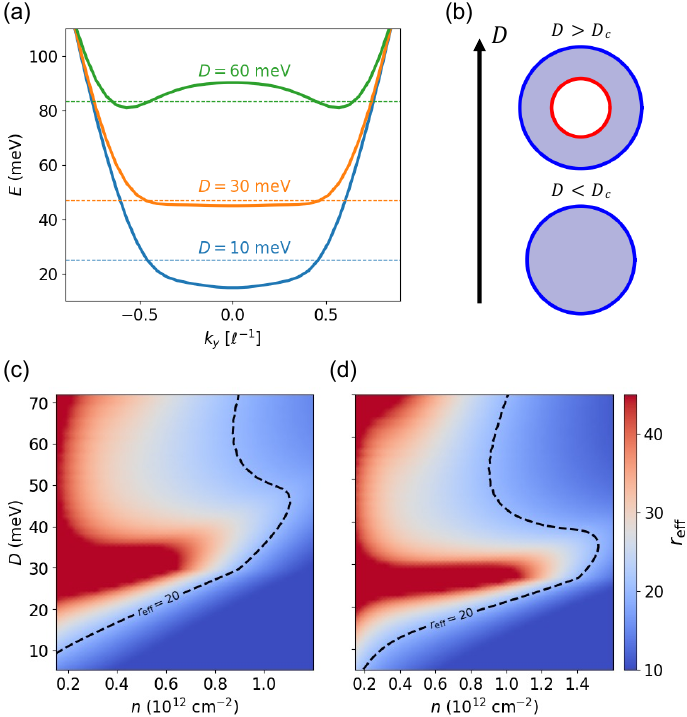}
    \caption{Microscopic band structure (a) and illustration of the simply-connected and annular Fermi sea regimes (blue shaded), with the black and red lines indicating electron- and hole-like Fermi surfaces, respectively. Ratio of interaction energy to kinetic energy in 4 (c) and 5 layer (d) rhombohedral graphene, with relative dielectric constant $\epsilon=5$. %The upper dashed line marks the onset of a hole pocket at $\k=0$ in the Mexican hat regime. \FG{This last sentence should be removed if we don't show the Lifshitz line.}
    }
    \label{fig:int_vs_kinetic}
\end{figure}

As shown in Fig.~\ref{fig:int_vs_kinetic}~(a), increasing $D$ progressively flattens the bottom of the conduction band until a critical value $D_c$, at which the band minimum shifts from $\k = 0$ (the Dirac point) to a finite momentum $|\k| = k_0$. %Across this evolution, 
Above $D_c$, the band is reshaped %from a monotonic dispersion with a single minimum at 
into a Mexican-hat dispersion. % with energy minima at finite momentum.  
Consequently, at fixed carrier density, the Fermi surface changes from circular to annular, as sketched in Fig.~\ref{fig:int_vs_kinetic}~(b).
At finite doping, the non-interacting Fermi surface topology also changes with the carrier density $n$. 
In particular, in the Mexican hat regime, as $n$ decreases the system changes from having a single electron pocket centered at $\k = 0$ to an annular Fermi sea with an outer electron-like Fermi surface and an inner hole-like Fermi surface. Additional details can be found in the Supplementary Materials (SM)~\cite{supplementary}. 
% Fermi surface consisting of multiple pockets. 

Our goal is to explore the consequence of the underlying band structure on the interacting phase diagram. 
As a first step, we introduce a dimensionless quantity 
% it is useful 
that measures the strength of Coulomb interactions relative to the kinetic energy in this system \cite{geier2024chiraltopologicalsuperconductivityisospin}. 
Specifically, we define %the dimensionless quantity %(the analog of $r_s$ parameter for parabolic band systems) 
\begin{equation}
r_{\rm eff} \equiv \frac{ E_{\rm int}}{E_{\rm kin}} = \dfrac{e^2/(4 \pi \epsilon\epsilon_0 d_e)}{\langle \varepsilon(\k) \rangle} ,    
\end{equation}
where $E_{\rm int}$ is the Coulomb energy evaluated at the average interparticle distance $d_e=1/\sqrt{\pi n}$, $E_{\rm int}=e^2/(4\pi\epsilon_0\epsilon d_e)$ and $\langle \varepsilon(\k) \rangle$ is the kinetic energy per-particle, averaged  over {\it all} occupied states in the non-interacting limit~\cite{supplementary}. 

For conventional two-dimensional electron gas with parabolic dispersion, $r_{\rm eff}$ reduces to the Wigner–Seitz radius $r_s$ determined by the carrier density ($r_{\rm eff} \propto n^{-1/2}$), and large $r_s$ at low density leads to the emergence of an electron Wigner crystal, driven by Coulomb interaction. In contrast, in rhombohedral graphene the ratio $r_{\rm eff}$ depends not only on the density, but also on the form of the underlying band dispersion tuned by the displacement field \cite{geier2024chiraltopologicalsuperconductivityisospin}.

Figure~\ref{fig:int_vs_kinetic}~(c) and~\ref{fig:int_vs_kinetic}~(d) show $r_{\rm eff} $ as a function of density $n$ and displacement field $D$ for four- and five-layer rhombohederal graphene, respectively, which allows us to identify the region of strong correlation. 
Remarkably, in the $(n,D)$ plane, the region of large $r_{\rm eff} \geq 20$---which corresponds to strongly correlated  liquid and Wigner crystals in the case of two-dimensional electron gas---matches well the region of experimentally observed insulating and superconducting states.   
In particular, $r_{\rm eff}$ generally evolves {\it non-monotonically} with the displacement field $D$ and the density $n$. 
%reflecting the strong field dependence of the conduction band dispersion relation. 
This trend mirrors the experimentally observed sequence in rhombohedral four-layer graphene~\cite{Han2025} in the density range $0.2-0.6\times 10^{12}\,\mathrm{cm}^{-2}$, where, at fixed $n$, the ground state evolves from a metal to a highly resistive state and then back to a metal as the displacement field is increased. 

It is instructive to compare the $r_{\rm eff}$ indicator---which takes into consideration all states below the Fermi level---with the Stoner criterion: $U D(E_F)>1$, where $D(E_F)$ is the noninteracting density of states {\it at} the Fermi level. In the Stoner picture, strong interaction effects are expected at the van-Hove singularity due to the divergent density of states. However, while the line of van-Hove singularity for rhombohedral graphene continues to very high density and displacement field \cite{supplementary}, the interaction-driven states are only observed below a certain density $\sim 1\times 10^{12}\,\mathrm{cm}^{-2}$, as correctly captured by $r_{\rm eff}$.

Our $r_{\rm eff}$ plot also correctly identifies strongly correlated region of $N=5, 6$ layer rhombohedral graphene. As shown in Fig.~\ref{fig:int_vs_kinetic}~(d), large $r_{\rm eff}$ appears at smaller displacement fields and extends over a broader range of electron densities compared to $N=4$~\cite{Han2025,han2026evidencemetallicwignercrystal}, which reflects the flatter band dispersion -- similar conclusions can be drawn for $N = 6$ layers, see the SM~\cite{supplementary}.
Thus, our indicator $r_{\rm eff}$ provides a useful guide for navigating the rich phase diagram of correlated electron states in rhombohedral graphene.

The discussion above highlights two key ingredients of rhombohedral multilayer graphene: the nonmonotonic evolution of the correlation strength and the topological transition of the Fermi surface from a single electron pocket to multiple pockets. In the following, we develop state-of-art neural network methods to solve many-body ground states variationally, focusing on the region of large $r_{\rm eff}$ and annular Fermi sea. %using neural-network variational Monte Carlo and Hartree Fock. 

\begin{figure*}
    \centering
    \label{fig:NNplots}
    \includegraphics[width=1.0\linewidth]{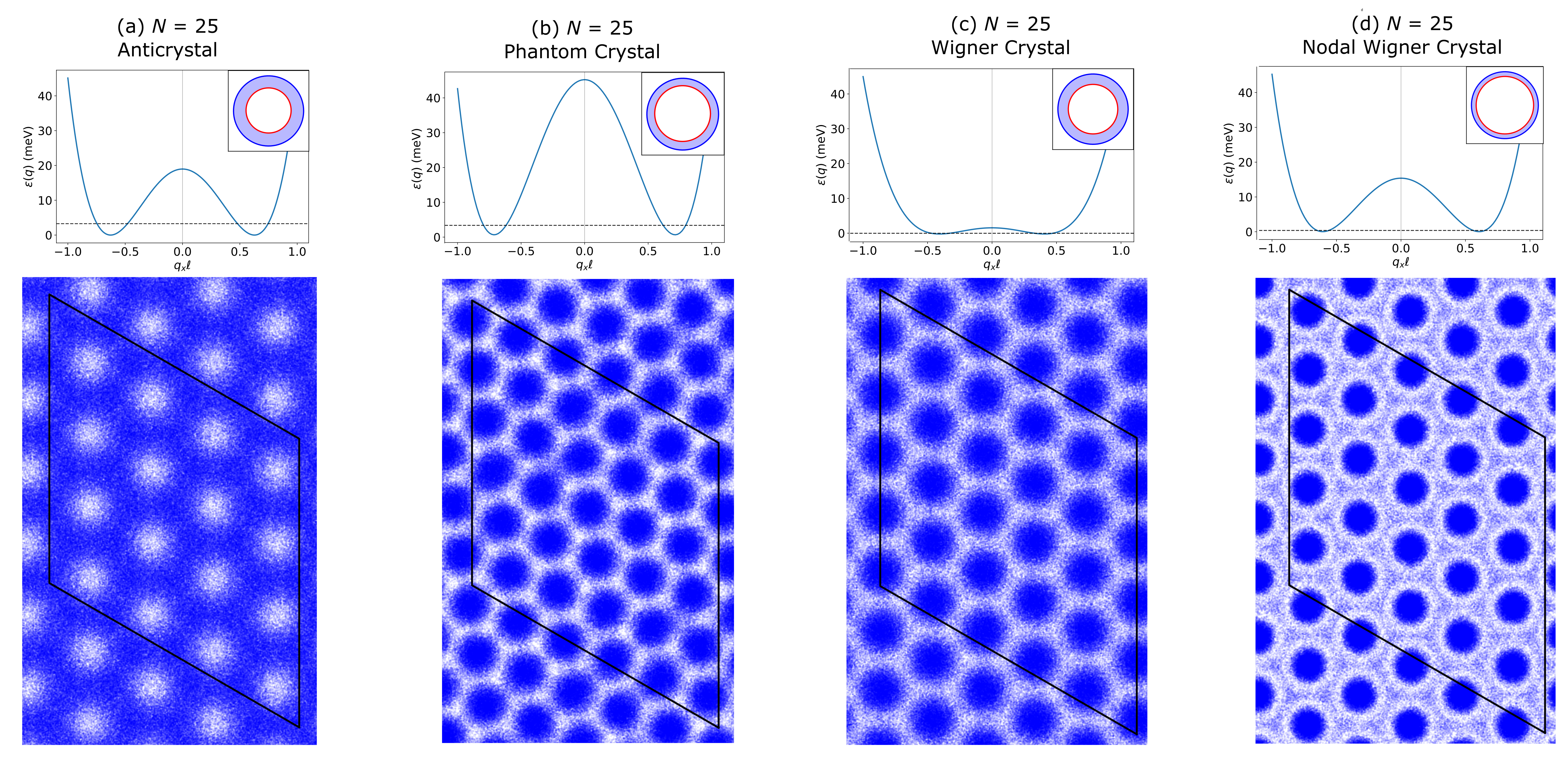}
    \caption{(a)--(d) Ground-state charge density obtained from NN-VMC, illustrating four representative crystalline states. Blue denotes regions of high electron density, while white denotes regions of low density. The parameters are $(\alpha/E_C, \gamma/E_C, n)=(0.582,0.740,0.86\times10^{12}~{\rm cm}^{-2})$ for (a), $(0.740,0.740,0.659\times10^{12}~{\rm cm}^{-2})$ for (b), $(0.232,0.702,0.293\times10^{12}~{\rm cm}^{-2})$ for (c), and $(0.544,0.739,0.293\times10^{12}~{\rm cm}^{-2})$ for (d). $E_C = e^2/ (4 \pi \epsilon \epsilon_0 \ell)$ is the Coulomb energy scale where $\ell = \hbar v_F/t_1$ is the characteristic length scale of graphene with the Fermi velocity $v_F$ and $t_1$ is inter-layer opposite sublattice hopping strength \cite{supplementary}.
    The parallelogram indicates the simulation supercell. Calculations are performed with relative dielectric constant $\epsilon=5$. Top panel in each plots shows the non-interaction dispersion: $\epsilon(\k) = -\alpha |\k|^2 + \gamma |\k|^4$ measured relative to the band minimum. The dashed line denotes the Fermi level.  }
    \label{fig:NNplots}
\end{figure*}

\par

\textit{Neural Network VMC ---}  In recent years, neural network variational Monte Carlo (NN-VMC) techniques have become an extremely powerful tool for strongly correlated quantum matter \cite{Carleo_2017, Carrasquilla_2017, Luo_2019, Pfau_2020}, enabling accurate variational studies of continuum electron gases \cite{Cassella_2023, Pescia_2024, Smith_2024, Valenti_2025, Gaggioli_2025, Gaggioli_2026}, superconducting states \cite{Li_2025}, fractionalized phases \cite{Teng_2025,Qian_2025, Nazaryan_2025, Fadon_2025,  Abouelkomsan_2026, Zhu_2026}, and other interacting many-body problems \cite{Kim_2024, Foster_2025, Linteau_2026, Abouelkomsan_2026_topological, Fu2026}. 
In particular, the ability of self-attention  architectures \cite{vonGlehn_2023, Geier_2025} to represent diverse quantum phases within a single expressive ansatz makes them well suited for rhombohedral graphene, where metallic, crystalline, and possibly other competing phases appear in close proximity.

Our self-attention--based wavefunction consists of a sum of $N_{\rm det}$ determinants of many-body orbitals, and is given by
\begin{equation}
\label{eq:ansatz}
    \Psi_{\{\theta\}}(\{\r_i\})  =  \sum_{n=1}^{N_{\rm det}} {\rm det}[\phi_j^{(n)}(\r_i; \{\r_{\neq i }\})]
\end{equation} 
where $\phi_j^n(\r_i; \{\r_{\neq i }\})$ is a generalized many-body orbital for the $i$-th particle, that also depends on the coordinates of the remaining particles $\r_{\neq i }$. $\{\theta\}$ denotes the variational parameters of the ansatz. The many-body orbitals are output entirely from a self-attention deep neural network \cite{vonGlehn_2023,Geier_2025}. 

The neural network parameters $\{\theta\}$ are optimized by minimizing the total energy $E = \langle \Psi | H | \Psi \rangle/\langle\Psi|\Psi \rangle$. For the NN-VMC, we focus on the role of the conduction band dispersion on interaction physics %ignoring the effects of underlying quantum geometry}. Furthermore, 
using a simplified dispersion: $\varepsilon(\k)=-\alpha |\k|^2 + \gamma |\k|^4$, which captures the transition from single pocket ($\alpha<0$) to Mexican hat ($\alpha>0$) regime. 
We stress that our NN architecture itself is completely system agnostic: the only input information about rhombohedral graphene lies in the loss function, which is the variational energy.   
Details of  NN-VMC implementation can be found in the Supplementary Material. 

Our NN-VMC discovers
four distinct types of crystalline states (Fig.~\ref{fig:NNplots}) in different regions of the $(n,D)$ phase diagram, where the band dispersion is Mexican hat and the noninteracting Fermi 
sea is annular.   
Particularly interesting are two unconventional quantum crystals  that occur at relatively \textit{high} electron densities.  %, in agreement with the broad extent of the correlated region in Fig.~\ref{fig:int_vs_kinetic}~(c) and (d). 
First, at lower displacement fields where the Mexican hat is more shallow, i.e., smaller values $k_0 = \sqrt{\alpha /2\gamma}$, we find a triangular lattice of density minima (holes) in an otherwise nearly uniform electronic background, as shown in Fig.~\ref{fig:NNplots}~(a). The period of this crystal is set by the density of holes, which occupy the top of the Mexican hat above the Fermi level.  %related to the wavevector associated with the inner hole-like  Fermi surface. 
We refer to this crystalline state as ``anticrystal''.  

The origin of the anticrystal phase can be qualitatively understood by considering the non-interacting dispersion and Fermi surfaces, shown in Fig.~\ref{fig:int_vs_kinetic}~(a) and (b).
Away from the band bottom at $|\k|=k_0$, the kinetic energy changes slowly around $\k = 0$, while increasing rapidly at large momenta. 
As a result, the electron- and hole pockets have very different masses such that, for sufficiently high electron densities, the heavier hole mass favors crystallization, while the lighter electrons prefer a liquid state.  

At higher displacement fields, i.e., in the deep Mexican hat regime, we instead find a self-doped crystal (Fig.~\ref{fig:NNplots}~(b)) in which the electron density is periodically modulated, forming a triangular crystal. However, this crystalline state contains more density peaks than the number of particles, for example $31$ peaks for $N=25$ particles. This means that the emergent crystal unit cell contains \textit{less} than one electron on average, distinguishing this state from a conventional Wigner crystal that has one localized electron per unit cell. Since the number of density maxima does not match the number of electrons, we refer to this state as a ``phantom'' crystal.

The appearance of the phantom crystal is also tied to the Mexican-hat dispersion: in addition to the mean interparticle spacing set by the density, the annular Fermi sea introduces another length scale set by $k_0^{-1}$. As a result, the dominant charge-ordering wave vector minimizing the total energy is set by the interplay of these two length scales. In our NN-VMC calculations, we find the magnitude of the charge-ordering wave vector $Q$ to compromise between the conventional Wigner-crystal wave vector set by the particle density and the characteristic wave vectors of the Mexican-hat dispersions that produce prominent features in the non-interacting static density-density susceptibility~\cite{supplementary}.
This is consistent with the fact that the self-doped crystal appears in an intermediate-coupling regime at high electron density, where neither weak-coupling fermiology nor strong-coupling arguments alone are quantitatively accurate. 

Moving to lower densities, we encounter two more types of electron crystals.
The first, realized in the shallow Mexican hat regime (smaller $D$), is the ordinary triangular Wigner crystal (WC), shown in Fig.~\ref{fig:NNplots}~(c), with one electron per unit cell. 
Moving to deeper Mexican hat dispersions (higher $D$), this state transforms into a $\textit{nodal}$ Wigner crystal (Fig.~\ref{fig:NNplots}~(d)). The nodal WC is characterized by a momentum space occupation $n(\k)$ that is concentrated around the non-interacting annular Fermi sea centered at $|\k|\approx k_0$ \cite{supplementary}.
In real space, the amplitude of the wavefunction is depleted along a ring centered around the maxima at each triangular lattice site. 

The peculiar shape of nodal crystal orbitals can be quantitatively understood by considering an effective single-particle problem with Mexican-hat dispersion, in which the particle moves in an approximately harmonic potential generated by interactions with neighboring localized electrons \cite{joy2023wigner, joy2025chiral, supplementary}.
When the momentum scale $k_0$ is large enough, the electron wavepacket is then built from plane waves with $|\k|\approx k_0$. The resulting ground state wavefunction consists of a Gaussian factor, typical of the harmonic oscillator ground state, times an oscillatory Bessel function that first vanishes along a ring with radius $\simeq 2.4048 / k_0$, in agreement with our numerical findings (cf SM \cite{supplementary}).  

The phantom crystal obtained in our calculations is consistent with recent Hartree-Fock results \cite{Dong_2026,Feng2026}. Additionally, our calculations predict the presence of an anticrystal at higher densities.

\begin{figure}
    \centering
    \includegraphics[width=\linewidth]{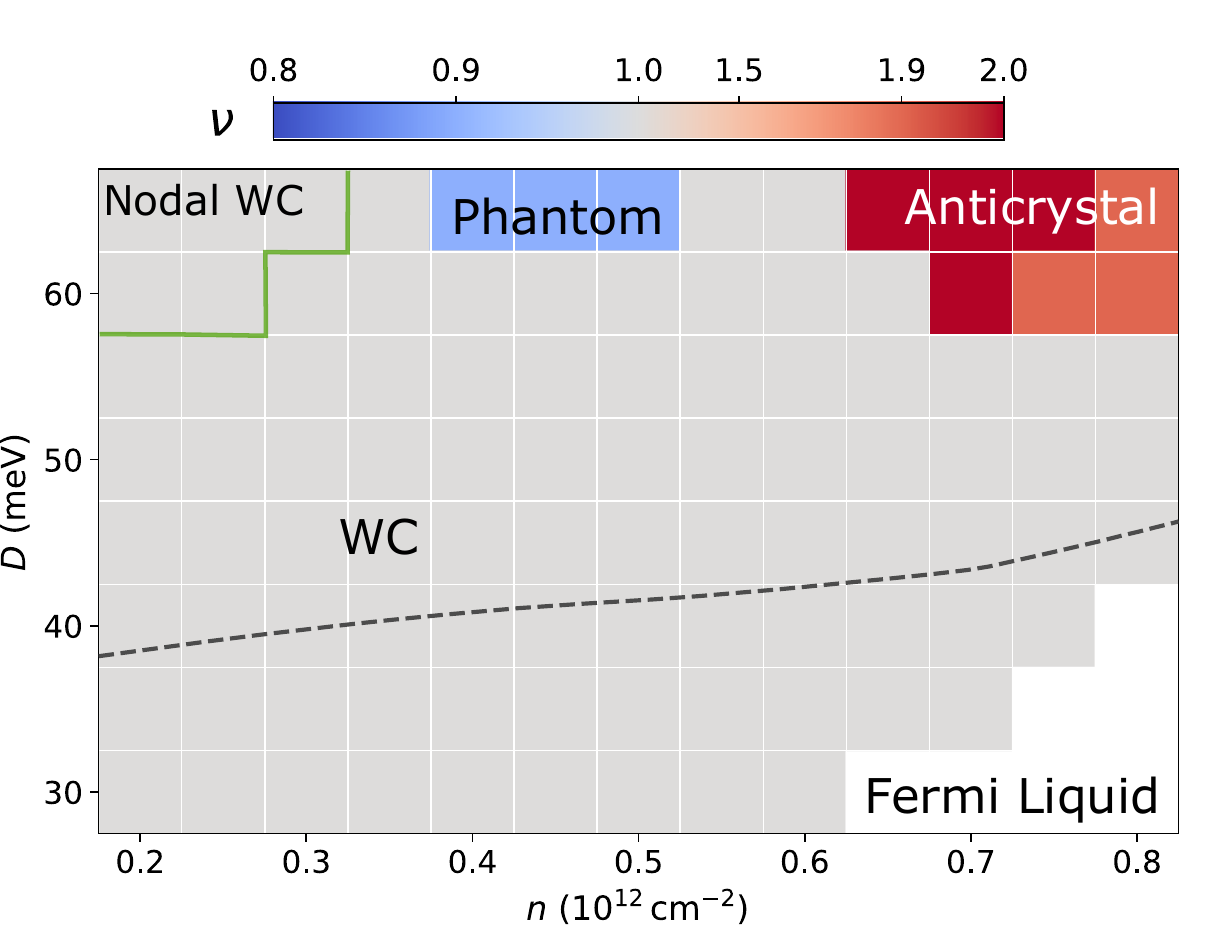}
    \caption{Hartree Fock phase diagram for electrons with dispersion $\epsilon(\k)$ of the conduction band of rhombohedral tetralayer graphene, interacting via the long-range Coulomb interaction with dielectric constant $\epsilon=12$. 
    The color code indicates the filling factor $\nu$ of the ground state crystal unit cell. The dashed line denotes the Lifshitz line where the Fermi surface topology chances.}
    
\label{fig:HF_phase_diagram}
\end{figure}

\textit{Hartree--Fock Phase Diagram ---}  Having presented our NN-VMC results on finite-size systems, we now provide complementary evidence for the observed crystalline states within the Hartree--Fock (HF) framework. 
To this end, we perform self-consistent calculations that minimize the Hartree--Fock energy for a prescribed real-space pattern directly in the thermodynamic limit, while further optimizing over the unit-cell lattice vectors $(\a_1,\a_2)$, which determine the filling factor $\nu=n|\a_1\times\a_2|$. In this optimization, we restrict our attention to triangular unit cells with a fixed aspect ratio of $1:1$ and consider a wide range of possible filling factors $\nu$.
Additional fully unrestricted HF calculations, performed for periodic boundary conditions at particular system sizes, are discussed in the SM \cite{supplementary}.

The filling factor $\nu$ of the crystal unit cell in the ground state is shown in Fig.~\ref{fig:HF_phase_diagram} as a function of the electron density and displacement field. 
Notably, our Hartree-Fock calculations based on the full band structure of rhombohedral graphene---which includes trigonal warping---reproduce all the unconventional crystalline states identified with our NN-VMC simulations.

\begin{figure}
    \centering
    \includegraphics[width=\linewidth]{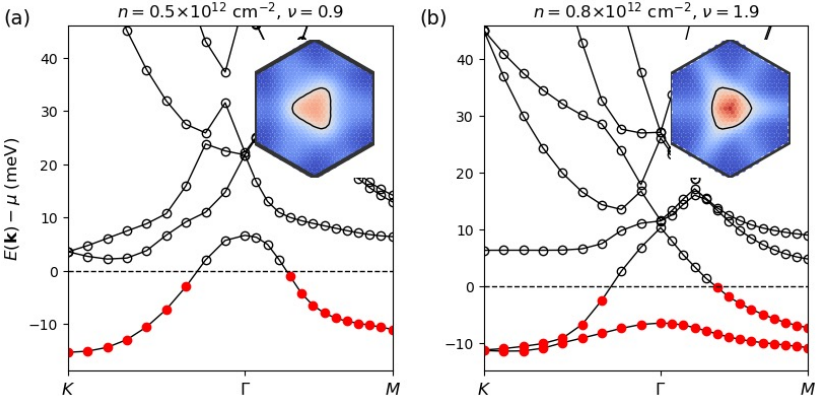}
    \caption{Hartree-Fock band structure of the self-doped Wigner crystal (a) and the metallic anticrystal (b).
    The inset of panels (a) and (b) show the Fermi surface of    hole-like quasiparticles. 
    Panel (a) is obtained at $n=0.5\times 10^{12}$cm$^{-2}$ and $D=65$meV ($\nu\approx0.9)$, and panel (b) at $n=0.8\times 10^{12}$cm$^{-2}$ and $D=60$meV ($\nu\approx1.9$), with dielectric constant $\epsilon=12$.}
    \label{fig:AC}
\end{figure}

In addition to the ordinary and nodal Wigner crystals, both of which have $\nu =1$ and therefore one electron per unit cell, the HF calculations reveal extended regions of the $(n,D)$ plane in which the optimal filling factor deviates from unity. At larger $D$ and intermediate densities, we find triangular phantom crystals with $\nu <1$. 
This metallic crystalline state can be understood as arising from hole doping of the insulating Wigner crystal at $\nu=1$, with a hole density equal to approximately $10\%$ of the commensurate electron density. 
The corresponding band structure and hole-like Fermi surface are shown in Fig.~\ref{fig:AC}~(a). 
Upon further increasing $n$, however, we find a distinct region in which the optimal filling factor is concentrated near $\nu =2$, corresponding to \textit{two} electrons per crystal unit cell. This region hosts the anticrystal phase, characterized by a triangular lattice of holes. 

In the anticrystal regime, the system first forms an \textit{insulating} Wigner crystal of holes with exact filling factor $\nu=2$. 
Upon changing the carrier density or the displacement field, this state evolves into the metallic anticrystal.
Unlike the phantom crystal, which emerges at incommensurate fillings near $\nu=1$, the metallic anticrystal originates in the vicinity of an insulating commensurate crystal with two electrons per unit cell.
Figure~\ref{fig:AC}~(b) shows the bands of the metallic anticrystal obtained at $n=0.8\times10^{12}\,\mathrm{cm}^{-2}$ and $D=60 \,\mathrm{meV}$.

For these parameters, we find the metallic anticrystal optimal doping to be $\nu\approx 1.9$ electrons per unit cell, corresponding to a hole doping of $\delta n\approx0.05\,n=0.04\times10^{12}\,\mathrm{cm}^{-2}$ relative to $\nu=2$. 
Thus, the system is hole-doped by approximately $5\%$ relative to the commensurate filling.
Interestingly, the hole quasiparticles form a Fermi surface around $\Gamma$, distorted by trigonal warping, as shown in the inset of Fig.~\ref{fig:AC}~(b).

\textit{Discussion ---} The interplay between strong electronic correlations and a tunable, nontrivial band structure makes rhombohedral graphene an exceptional platform for exploring strongly correlated quantum matter. In this work, we advance the understanding of this fascinating material by introducing the dimensionless ratio $r_{\rm eff}$ as a useful guide for navigating its complex phase diagram. Combining state-of-the-art NN-VMC with Hartree–Fock calculations, we uncover a rich variety of crystalline states enabled by the tunable fermiology of rhombohedral graphene. 

In particular, the anticrystal and self-doped crystal phases emerge from the competition among the distinct microscopic length scales associated with the annular Fermi-sea regime, and could be distinguished experimentally through direct imaging using scanning tunneling microscopy~\cite{Li_2021,Tsui_2024}. 
Furthermore, our results are consistent with recent experimental reports of metallic crystals~\cite{han2026evidencemetallicwignercrystal,zhou2026competingordersdrivenwigner} with $15\%$ hole doping at electron densities $n\sim 0.4\times 10^{12}\,$cm$^{-2}$, which we identify as the self-doped Wigner crystal.  

More importantly, we highlight the resistive state observed at higher density $n\sim 0.6-0.8\times 10^{12}\,$cm$^{-2}$, which intersects two chiral superconductivity domes at lower and higher displacement fields respectively. Our work identifies this resistive state, which is located in a region with large $r_{\rm eff}$ \cite{geier2024chiraltopologicalsuperconductivityisospin}, as an anticrystal. Given its close proximity to superconductivity, the interplay of crystalline order and superconductivity may hold the key for a pairing mechanism in rhombohedral graphene.

An important direction for future work is to study the effects of band topology on crystallization. In the presence of finite Berry curvature, crystalline order may also coexist with both quantized or unquantized anomalous Hall response \cite{sheng2024quantum,Dong_2024,dong2024theory, zhou2024fractional,Valenti_2025,soejima2025lambda, tan2024parent, desrochers2026elastic,zeng2024,soejima2024anomalous,dong2024stability}. 
Incorporating these Berry curvature effects beyond band dispersion in the strongly correlated regime of rhombohedral graphene is an important subject for further study.

\textit{Note Added ---} A related study of crystalline phases in rhombohedral graphene will appear in a forthcoming preprint~\cite{LM2608}. 

\textit{Acknowledgments ---} 
We acknowledge helpful discussions with Long Ju, Tonghang Han, Max Geier, Margarita Davydova,  Luke Kim and Andrea Young. We thank Xiaomeng Liu for interesting and fruitful discussions. 
This work was supported by a Simons Investigator Award from the Simons Foundation. 
We acknowledge MIT SuperCloud
and the Lincoln Laboratory Supercomputing Center, IAIFI (the NSF Cooperative Agreement PHY-2019786) and MIT Engaging cluster for providing the computational resources that have contributed to the results reported in this paper.   

\bibliography{biblio}

@article{Tsui_2024,
	author = {Tsui, Yen-Chen and He, Minhao and Hu, Yuwen and Lake, Ethan and Wang, Taige and Watanabe, Kenji and Taniguchi, Takashi and Zaletel, Michael P. and Yazdani, Ali},
	date = {2024/04/01},
	doi = {10.1038/s41586-024-07212-7},
	id = {Tsui2024},
	isbn = {1476-4687},
	journal = {Nature},
	number = {8007},
	pages = {287--292},
	title = {Direct observation of a magnetic-field-induced Wigner crystal},
	url = {https://doi.org/10.1038/s41586-024-07212-7},
	volume = {628},
	year = {2024}}

@article{Li_2021,
	author = {Li, Hongyuan and Li, Shaowei and Regan, Emma C. and Wang, Danqing and Zhao, Wenyu and Kahn, Salman and Yumigeta, Kentaro and Blei, Mark and Taniguchi, Takashi and Watanabe, Kenji and Tongay, Sefaattin and Zettl, Alex and Crommie, Michael F. and Wang, Feng},
	date = {2021/09/01},
	doi = {10.1038/s41586-021-03874-9},
	id = {Li2021},
	isbn = {1476-4687},
	journal = {Nature},
	number = {7878},
	pages = {650--654},
	title = {Imaging two-dimensional generalized Wigner crystals},
	url = {https://doi.org/10.1038/s41586-021-03874-9},
	volume = {597},
	year = {2021}}

@article{Koshino_2009,
  title = {Trigonal warping and Berry's phase $N\ensuremath{\pi}$ in ABC-stacked multilayer graphene},
  author = {Koshino, Mikito and McCann, Edward},
  journal = {Phys. Rev. B},
  volume = {80},
  issue = {16},
  pages = {165409},
  numpages = {8},
  year = {2009},
  month = {Oct},
  publisher = {American Physical Society},
  doi = {10.1103/PhysRevB.80.165409},
  url = {https://link.aps.org/doi/10.1103/PhysRevB.80.165409}
}

@article{Zhang_MacDonald_2010,
  title = {Band structure of $ABC$-stacked graphene trilayers},
  author = {Zhang, Fan and Sahu, Bhagawan and Min, Hongki and MacDonald, A. H.},
  journal = {Phys. Rev. B},
  volume = {82},
  issue = {3},
  pages = {035409},
  numpages = {10},
  year = {2010},
  month = {Jul},
  publisher = {American Physical Society},
  doi = {10.1103/PhysRevB.82.035409},
  url = {https://link.aps.org/doi/10.1103/PhysRevB.82.035409}
}

@article{Abouelkomsan_2026_topological,
   title={Topological order in neural wavefunctions},
   volume={113},
   ISSN={2469-9969},
   url={http://dx.doi.org/10.1103/bzq1-123h},
   DOI={10.1103/bzq1-123h},
   number={20},
   journal={Physical Review B},
   publisher={American Physical Society (APS)},
   author={Abouelkomsan, Ahmed and Geier, Max and Fu, Liang},
   year={2026},
   month=May }

@article{Carrasquilla_2017,
	author = {Carrasquilla, Juan and Melko, Roger G.},
	date = {2017/05/01},
	doi = {10.1038/nphys4035},
	id = {Carrasquilla2017},
	isbn = {1745-2481},
	journal = {Nature Physics},
	number = {5},
	pages = {431--434},
	title = {Machine learning phases of matter},
	url = {https://doi.org/10.1038/nphys4035},
	volume = {13},
	year = {2017}}

@article{Pfau_2020,
  title = {Ab initio solution of the many-electron Schr\"odinger equation with deep neural networks},
  author = {Pfau, David and Spencer, James S. and Matthews, Alexander G. D. G. and Foulkes, W. M. C.},
  journal = {Phys. Rev. Res.},
  volume = {2},
  issue = {3},
  pages = {033429},
  numpages = {20},
  year = {2020},
  month = {Sep},
  publisher = {American Physical Society},
  doi = {10.1103/PhysRevResearch.2.033429},
  url = {https://link.aps.org/doi/10.1103/PhysRevResearch.2.033429}
}

@article{Luo_2019,
  title = {Backflow Transformations via Neural Networks for Quantum Many-Body Wave Functions},
  author = {Luo, Di and Clark, Bryan K.},
  journal = {Phys. Rev. Lett.},
  volume = {122},
  issue = {22},
  pages = {226401},
  numpages = {6},
  year = {2019},
  month = {Jun},
  publisher = {American Physical Society},
  doi = {10.1103/PhysRevLett.122.226401},
  url = {https://link.aps.org/doi/10.1103/PhysRevLett.122.226401}
}

@article{Carleo_2017,
	author = {Giuseppe Carleo and Matthias Troyer},
	doi = {10.1126/science.aag2302},
	eprint = {https://www.science.org/doi/pdf/10.1126/science.aag2302},
	journal = {Science},
	number = {6325},
	pages = {602-606},
	title = {Solving the quantum many-body problem with artificial neural networks},
	url = {https://www.science.org/doi/abs/10.1126/science.aag2302},
	volume = {355},
	year = {2017}}

@article{Kim_2024,
	author = {Kim, Jane and Pescia, Gabriel and Fore, Bryce and Nys, Jannes and Carleo, Giuseppe and Gandolfi, Stefano and Hjorth-Jensen, Morten and Lovato, Alessandro},
	date = {2024/05/08},
	doi = {10.1038/s42005-024-01613-w},
	id = {Kim2024},
	isbn = {2399-3650},
	journal = {Communications Physics},
	number = {1},
	pages = {148},
	title = {Neural-network quantum states for ultra-cold Fermi gases},
	url = {https://doi.org/10.1038/s42005-024-01613-w},
	volume = {7},
	year = {2024}}

@misc{Foster_2025,
      title={An ab initio foundation model of wavefunctions that accurately describes chemical bond breaking}, 
      author={Adam Foster and Zeno Schätzle and P. Bernát Szabó and Lixue Cheng and Jonas Köhler and Gino Cassella and Nicholas Gao and Jiawei Li and Frank Noé and Jan Hermann},
      year={2025},
      eprint={2506.19960},
      archivePrefix={arXiv},
      primaryClass={physics.chem-ph},
      url={https://arxiv.org/abs/2506.19960}, 
}

@article{Pescia_2024,
  title = {Message-passing neural quantum states for the homogeneous electron gas},
  author = {Pescia, Gabriel and Nys, Jannes and Kim, Jane and Lovato, Alessandro and Carleo, Giuseppe},
  journal = {Phys. Rev. B},
  volume = {110},
  issue = {3},
  pages = {035108},
  numpages = {11},
  year = {2024},
  month = {Jul},
  publisher = {American Physical Society},
  doi = {10.1103/PhysRevB.110.035108},
  url = {https://link.aps.org/doi/10.1103/PhysRevB.110.035108}
}

@misc{Gaggioli_2026,
      title={Accurate Self-Attention Wavefunctions at Large Scale}, 
      author={Filippo Gaggioli and Sam Azadi and Liang Fu},
      year={2026},
      eprint={2607.08616},
      archivePrefix={arXiv},
      primaryClass={cond-mat.str-el},
      url={https://arxiv.org/abs/2607.08616}, 
}

@misc{Zhu_2026,
      title={Crystallization in the Fractional Quantum Hall Regime with Disorder-Aware Neural Quantum States}, 
      author={Jihang Zhu and Yi Huang and Xiaodong Hu and Di Xiao and Ting Cao},
      year={2026},
      eprint={2604.06316},
      archivePrefix={arXiv},
      primaryClass={cond-mat.str-el},
      url={https://arxiv.org/abs/2604.06316}, 
}

@misc{Fadon_2025,
      title={Extracting Anyon Statistics from Neural Network Fractional Quantum Hall States}, 
      author={Andres Perez Fadon and David Pfau and James S. Spencer and Wan Tong Lou and Titus Neupert and W. M. C. Foulkes},
      year={2025},
      eprint={2512.15872},
      archivePrefix={arXiv},
      primaryClass={cond-mat.str-el},
      url={https://arxiv.org/abs/2512.15872}, 
}

@article{Teng_2025,
  title = {Solving the fractional quantum Hall problem with self-attention neural network},
  author = {Teng, Yi and Dai, David D. and Fu, Liang},
  journal = {Phys. Rev. B},
  volume = {111},
  issue = {20},
  pages = {205117},
  numpages = {10},
  year = {2025},
  month = {May},
  publisher = {American Physical Society},
  doi = {10.1103/PhysRevB.111.205117},
  url = {https://link.aps.org/doi/10.1103/PhysRevB.111.205117}
}

@misc{Nazaryan_2025,
      title={Artificial Intelligence for Quantum Matter: Finding a Needle in a Haystack}, 
      author={Khachatur Nazaryan and Filippo Gaggioli and Yi Teng and Liang Fu},
      year={2025},
      eprint={2507.13322},
      archivePrefix={arXiv},
      primaryClass={cond-mat.str-el},
      url={https://arxiv.org/abs/2507.13322}, 
}

@misc{Li_2025,
      title={Attention is all you need to solve chiral superconductivity}, 
      author={Chun-Tse Li and Tzen Ong and Max Geier and Hsin Lin and Liang Fu},
      year={2025},
      eprint={2509.03683},
      archivePrefix={arXiv},
      primaryClass={cond-mat.supr-con},
      url={https://arxiv.org/abs/2509.03683}, 
}

@misc{Gaggioli_2025,
  title = {Electronic Crystals and Quasicrystals in Semiconductor Quantum Wells: An {{AI-powered}} Discovery},
  shorttitle = {Electronic Crystals and Quasicrystals in Semiconductor Quantum Wells},
  author = {Gaggioli, Filippo and Graham, Pierre-Antoine and Fu, Liang},
  year = 2025,
  month = dec,
  number = {arXiv:2512.10909},
  eprint = {2512.10909},
  primaryclass = {cond-mat},
  publisher = {arXiv},
  doi = {10.48550/arXiv.2512.10909},
}

@misc{Valenti_2025,
      title={Quantum Geometry Driven Crystallization: A Neural-Network Variational Monte Carlo Study}, 
      author={Agnes Valenti and Yaar Vituri and Yubo Yang and Daniel E. Parker and Tomohiro Soejima and Junkai Dong and Miguel A. Morales and Ashvin Vishwanath and Erez Berg and Shiwei Zhang},
      year={2025},
      eprint={2512.07947},
      archivePrefix={arXiv},
      primaryClass={cond-mat.str-el},
      url={https://arxiv.org/abs/2512.07947}, 
}

@article{soejima2025lambda,
  title={lambda-Jellium Model for the Anomalous Hall Crystal},
  author={Soejima, Tomohiro and Dong, Junkai and Vishwanath, Ashvin and Parker, Daniel E},
  journal={arXiv preprint arXiv:2503.12704},
  year={2025}
}

@article{sheng2024quantum,
  title={Quantum anomalous Hall crystal at fractional filling of moir{\'e} superlattices},
  author={Sheng, DN and Reddy, Aidan P and Abouelkomsan, Ahmed and Bergholtz, Emil J and Fu, Liang},
  journal={Physical Review Letters},
  volume={133},
  number={6},
  pages={066601},
  year={2024},
  publisher={APS}
}

@article{zeng2024,
  title = {Sublattice Structure and Topology in Spontaneously Crystallized Electronic States},
  author = {Zeng, Yongxin and Guerci, Daniele and Cr\'epel, Valentin and Millis, Andrew J. and Cano, Jennifer},
  journal = {Phys. Rev. Lett.},
  volume = {132},
  issue = {23},
  pages = {236601},
  numpages = {8},
  year = {2024},
  month = {Jun},
  publisher = {American Physical Society},
  doi = {10.1103/PhysRevLett.132.236601},
  url = {https://link.aps.org/doi/10.1103/PhysRevLett.132.236601}
}

@article{Smith_2024,
  title = {Unified Variational Approach Description of Ground-State Phases of the Two-Dimensional Electron Gas},
  author = {Smith, Conor and Chen, Yixiao and Levy, Ryan and Yang, Yubo and Morales, Miguel A. and Zhang, Shiwei},
  journal = {Phys. Rev. Lett.},
  volume = {133},
  issue = {26},
  pages = {266504},
  numpages = {6},
  year = {2024},
  month = {Dec},
  publisher = {American Physical Society},
  doi = {10.1103/PhysRevLett.133.266504},
  url = {https://link.aps.org/doi/10.1103/PhysRevLett.133.266504}
}

@article{Cassella_2023,
  title = {Discovering Quantum Phase Transitions with Fermionic Neural Networks},
  author = {Cassella, Gino and Sutterud, Halvard and Azadi, Sam and Drummond, N. D. and Pfau, David and Spencer, James S. and Foulkes, W. M. C.},
  journal = {Phys. Rev. Lett.},
  volume = {130},
  issue = {3},
  pages = {036401},
  numpages = {6},
  year = {2023},
  month = {Jan},
  publisher = {American Physical Society},
  doi = {10.1103/PhysRevLett.130.036401},
  url = {https://link.aps.org/doi/10.1103/PhysRevLett.130.036401}
}

@article{Geier_2025,
  title={Self-attention neural network for solving correlated electron problems in solids},
  author={Geier, Max and Nazaryan, Khachatur and Zaklama, Timothy and Fu, Liang},
  journal={Physical Review B},
  volume={112},
  number={4},
  pages={045119},
  year={2025},
  url={https://doi.org/10.1103/qxc3-bkc7}, 
  publisher={APS}
}

@misc{vonGlehn_2023,
      title={A Self-Attention Ansatz for Ab-initio Quantum Chemistry}, 
      author={Ingrid von Glehn and James S. Spencer and David Pfau},
      year={2023},
      eprint={2211.13672},
      archivePrefix={arXiv},
      primaryClass={physics.chem-ph},
      url={https://arxiv.org/abs/2211.13672}, 
}

@misc{Abouelkomsan_2026,
      title={First-Principles AI finds crystallization of fractional quantum Hall liquids}, 
      author={Ahmed Abouelkomsan and Liang Fu},
      year={2026},
      eprint={2602.03927},
      archivePrefix={arXiv},
      primaryClass={cond-mat.mes-hall},
      url={https://arxiv.org/abs/2602.03927}, 
}

@article{Linteau_2026,
   title={Neural wave functions for high-pressure atomic hydrogen},
   volume={8},
   ISSN={2643-1564},
   url={http://dx.doi.org/10.1103/t72h-tkcx},
   DOI={10.1103/t72h-tkcx},
   number={2},
   journal={Physical Review Research},
   publisher={American Physical Society (APS)},
   author={Linteau, David and Moroni, Saverio and Carleo, Giuseppe and Holzmann, Markus},
   year={2026},
   month=Apr }

@article{Qian_2025,
  title = {Describing Landau Level Mixing in Fractional Quantum Hall States with Deep Learning},
  author = {Qian, Yubing and Zhao, Tongzhou and Zhang, Jianxiao and Xiang, Tao and Li, Xiang and Chen, Ji},
  journal = {Phys. Rev. Lett.},
  volume = {134},
  issue = {17},
  pages = {176503},
  numpages = {8},
  year = {2025},
  month = {Apr},
  publisher = {American Physical Society},
  doi = {10.1103/PhysRevLett.134.176503},
  url = {https://link.aps.org/doi/10.1103/PhysRevLett.134.176503}
}

@misc{Fu2026,
      title={Fermi Sets: Universal and interpretable neural architectures for fermions}, 
      author={Liang Fu},
      year={2026},
      eprint={2601.02508},
      archivePrefix={arXiv},
      primaryClass={cond-mat.str-el},
      url={https://arxiv.org/abs/2601.02508}, 
}

@article{tan2024parent,
  title={Parent Berry curvature and the ideal anomalous Hall crystal},
  author={Tan, Tixuan and Devakul, Trithep},
  journal={Physical Review X},
  volume={14},
  number={4},
  pages={041040},
  year={2024},
  publisher={APS}
}

@article{desrochers2026elastic,
  title={Elastic response and instabilities of anomalous Hall crystals},
  author={Desrochers, F{\'e}lix and Hirsbrunner, Mark R and Huxford, Joe and Patri, Adarsh S and Senthil, T and Kim, Yong Baek},
  journal={Physical Review Letters},
  volume={136},
  number={16},
  pages={166503},
  year={2026},
  publisher={APS}
}

@article{soejima2024anomalous,
  title={Anomalous Hall crystals in rhombohedral multilayer graphene. II. General mechanism and a minimal model},
  author={Soejima, Tomohiro and Dong, Junkai and Wang, Taige and Wang, Tianle and Zaletel, Michael P and Vishwanath, Ashvin and Parker, Daniel E},
  journal={Physical Review B},
  volume={110},
  number={20},
  pages={205124},
  year={2024},
  publisher={APS}
}

@article{dong2024stability,
  title={Stability of anomalous Hall crystals in multilayer rhombohedral graphene},
  author={Dong, Zhihuan and Patri, Adarsh S and Senthil, T},
  journal={Physical Review B},
  volume={110},
  number={20},
  pages={205130},
  year={2024},
  publisher={APS}
}

@misc{Dong_2026,
      title={Crystals Caught Doping: Metallic Wigner Crystals in Rhombohedral Graphene}, 
      author={Junkai Dong and Tomohiro Soejima and Daniel E. Parker and Ashvin Vishwanath},
      year={2026},
      eprint={2604.00114},
      archivePrefix={arXiv},
      primaryClass={cond-mat.str-el},
      url={https://arxiv.org/abs/2604.00114}, 
}

@misc{Feng2026,
      title={Self-doped Crystal from Preempted Band-inversion Transitions}, 
      author={Jiechao Feng and Zhaoyu Han and Michael P. Zaletel and Zhihuan Dong},
      year={2026},
      eprint={2604.09820},
      archivePrefix={arXiv},
      primaryClass={cond-mat.str-el},
      url={https://arxiv.org/abs/2604.09820}, 
}

@article{joy2023wigner,
  title={Wigner crystallization in Bernal bilayer graphene},
  author={Joy, Sandeep and Skinner, Brian},
  journal={arXiv preprint arXiv:2310.07751},
  year={2023}
}

@article{joy2025chiral,
  title={Chiral Wigner crystal phases induced by Berry curvature},
  author={Joy, Sandeep and Levitov, Leonid and Skinner, Brian},
  journal={Physical Review Letters},
  volume={135},
  number={25},
  pages={256502},
  year={2025},
  publisher={APS}
}

@article{Dong_2024,
   title={Anomalous Hall Crystals in Rhombohedral Multilayer Graphene. I. Interaction-Driven Chern Bands and Fractional Quantum Hall States at Zero Magnetic Field},
   volume={133},
   ISSN={1079-7114},
   url={http://dx.doi.org/10.1103/PhysRevLett.133.206503},
   DOI={10.1103/physrevlett.133.206503},
   number={20},
   journal={Physical Review Letters},
   publisher={American Physical Society (APS)},
   author={Dong, Junkai and Wang, Taige and Wang, Tianle and Soejima, Tomohiro and Zaletel, Michael P. and Vishwanath, Ashvin and Parker, Daniel E.},
   year={2024},
   month=nov }

@article{Han2025,
   title={Signatures of chiral superconductivity in rhombohedral graphene},
   volume={643},
   ISSN={1476-4687},
   url={http://dx.doi.org/10.1038/s41586-025-09169-7},
   DOI={10.1038/s41586-025-09169-7},
   number={8072},
   journal={Nature},
   publisher={Springer Science and Business Media LLC},
   author={Han, Tonghang and Lu, Zhengguang and Hadjri, Zach and Shi, Lihan and Wu, Zhenghan and Xu, Wei and Yao, Yuxuan and Cotten, Armel A. and Sharifi Sedeh, Omid and Weldeyesus, Henok and Yang, Jixiang and Seo, Junseok and Ye, Shenyong and Zhou, Muyang and Liu, Haoyang and Shi, Gang and Hua, Zhenqi and Watanabe, Kenji and Taniguchi, Takashi and Xiong, Peng and Zumbühl, Dominik M. and Fu, Liang and Ju, Long},
   year={2025},
   month=May, pages={654–661} }

@misc{han2026evidencemetallicwignercrystal,
      title={Evidence of Metallic Wigner Crystal in Rhombohedral Graphene}, 
      author={Tonghang Han and Jackson P. Butler and Shenyong Ye and Zhenqi Hua and Surajit Dutta and Zach Hadjri and Zhenghan Wu and Jixiang Yang and Junseok Seo and Phatthanon Pattanakanvijit and Emily Aitken and Kenji Watanabe and Takashi Taniguchi and Peng Xiong and Eli Zeldov and Zhengguang Lu and Raymond Ashoori and Long Ju},
      year={2026},
      eprint={2604.00113},
      archivePrefix={arXiv},
      primaryClass={cond-mat.mes-hall},
      url={https://arxiv.org/abs/2604.00113}, 
}

@misc{supplementary,
	Note = {See the Supplemental Material [URL] for details...}}

@misc{LM2608,
    title= {Shape of Wigner Crystal and Hole Self-Doping in a Mexican-Hat Dispersion},
    author = {Kim, Minho Luke and Wen, Xiao-Gang},
	eprint = {to appear}}

@article{han2024correlated,
  title={Correlated insulator and Chern insulators in pentalayer rhombohedral-stacked graphene},
  author={Han, Tonghang and Lu, Zhengguang and Scuri, Giovanni and Sung, Jiho and Wang, Jue and Han, Tianyi and Watanabe, Kenji and Taniguchi, Takashi and Park, Hongkun and Ju, Long},
  journal={Nature Nanotechnology},
  volume={19},
  number={2},
  pages={181--187},
  year={2024},
  publisher={Nature Publishing Group UK London}, 
url= {https://www.nature.com/articles/s41565-023-01520-1.pdf}
}

@article{han2025signatures,
  title={Signatures of chiral superconductivity in rhombohedral graphene},
  author={Han, Tonghang and Lu, Zhengguang and Hadjri, Zach and Shi, Lihan and Wu, Zhenghan and Xu, Wei and Yao, Yuxuan and Cotten, Armel A and Sharifi Sedeh, Omid and Weldeyesus, Henok and others},
  journal={Nature},
  volume={643},
  number={8072},
  pages={654--661},
  year={2025},
  publisher={Nature Publishing Group UK London}
}

@misc{dutta2026reconfigurablechiralsuperconductivity,
      title={Reconfigurable chiral superconductivity}, 
      author={Surajit Dutta and Nadav Auerbach and Tonghang Han and Yaozhang Zhou and Gal Shavit and Niladri-Sekhar Kander and Yuri Myasoedov and Martin E. Huber and Kenji Watanabe and Takashi Taniguchi and Long Ju and Eli Zeldov},
      year={2026},
      eprint={2605.13303},
      archivePrefix={arXiv},
      primaryClass={cond-mat.mes-hall},
      url={https://arxiv.org/abs/2605.13303}, 
}

@article{dong2024theory,
  title={Theory of quantum anomalous Hall phases in pentalayer rhombohedral graphene moir{\'e} structures},
  author={Dong, Zhihuan and Patri, Adarsh S and Senthil, Todadri},
  journal={Physical Review Letters},
  volume={133},
  number={20},
  pages={206502},
  year={2024},
  publisher={APS}
}

@article{zhou2024fractional,
  title={Fractional quantum anomalous Hall effect in rhombohedral multilayer graphene in the moir{\'e}less limit},
  author={Zhou, Boran and Yang, Hui and Zhang, Ya-Hui},
  journal={Physical Review Letters},
  volume={133},
  number={20},
  pages={206504},
  year={2024},
  publisher={APS}
}

@article{Lu2024Feb,
	author = {Lu, Zhengguang and Han, Tonghang and Yao, Yuxuan and Reddy, Aidan P. and Yang, Jixiang and Seo, Junseok and Watanabe, Kenji and Taniguchi, Takashi and Fu, Liang and Ju, Long},
	title = {{Fractional quantum anomalous Hall effect in multilayer graphene}},
	journal = {Nature},
	volume = {626},
	pages = {759--764},
	year = {2024},
	month = feb,
	issn = {1476-4687},
	publisher = {Nature Publishing Group},
	doi = {10.1038/s41586-023-07010-7}
}

@article{lu_extended_2025,
	title = {Extended quantum anomalous {Hall} states in graphene/{hBN} moiré superlattices},
	volume = {637},
	issn = {1476-4687},
	url = {https://doi.org/10.1038/s41586-024-08470-1},
	doi = {10.1038/s41586-024-08470-1},
	number = {8048},
	journal = {Nature},
	author = {Lu, Zhengguang and Han, Tonghang and Yao, Yuxuan and Hadjri, Zach and Yang, Jixiang and Seo, Junseok and Shi, Lihan and Ye, Shenyong and Watanabe, Kenji and Taniguchi, Takashi and Ju, Long},
	month = jan,
	year = {2025},
	pages = {1090--1095},
}

@misc{xie2025tunablefractionalcherninsulators,
      title={Tunable Fractional Chern Insulators in Rhombohedral Graphene Superlattices}, 
      author={Jian Xie and Zihao Huo and Xin Lu and Zuo Feng and Zaizhe Zhang and Wenxuan Wang and Qiu Yang and Kenji Watanabe and Takashi Taniguchi and Kaihui Liu and Zhida Song and X. C. Xie and Jianpeng Liu and Xiaobo Lu},
      year={2025},
      eprint={2405.16944},
      archivePrefix={arXiv},
      primaryClass={cond-mat.mes-hall},
      url={https://arxiv.org/abs/2405.16944}, 
}

@misc{nguyen2025hierarchysuperconductivitytopologicalcharge,
      title={A Hierarchy of Superconductivity and Topological Charge Density Wave States in Rhombohedral Graphene}, 
      author={Ron Q. Nguyen and Hai-Tian Wu and Erin Morissette and Naiyuan J. Zhang and Peiyu Qin and Kenji Watanabe and Takashi Taniguchi and Aaron W. Hui and Dima E. Feldman and J. I. A. Li},
      year={2025},
      eprint={2507.22026},
      archivePrefix={arXiv},
      primaryClass={cond-mat.mes-hall},
      url={https://arxiv.org/abs/2507.22026}, 
}

@misc{zhou2026competingordersdrivenwigner,
      title={Competing Orders Driven by Wigner Crystal Phase in Rhombohedral Graphene}, 
      author={Zekang Zhou and Kilian Krötzsch and Raphaël Ayache and Yonggen Li and Sandeep Joy and Kenji Watanabe and Takashi Taniguchi and Moty Heiblum and Preden Roulleau and Mitali Banerjee},
      year={2026},
      eprint={2607.15014},
      archivePrefix={arXiv},
      primaryClass={cond-mat.mes-hall},
      url={https://arxiv.org/abs/2607.15014}, 
}

@misc{qin2026stripeordermetallicsuperconducting,
      title={Stripe Order in the Metallic and Superconducting Phases of Rhombohedral Hexalayer Graphene}, 
      author={Peiyu Qin and Hai-Tian Wu and Ron Q. Nguyen and Erin Morissette and Naiyuan J. Zhang and K. Watanabe and T. Taniguchi and J. I. A. Li},
      year={2026},
      eprint={2504.05129},
      archivePrefix={arXiv},
      primaryClass={cond-mat.mes-hall},
      url={https://arxiv.org/abs/2504.05129}, 
}

@misc{nguyen2026coexistingchargedensitywave,
      title={Coexisting Charge Density Wave and Superconducting Order in Quantizing Magnetic Fields}, 
      author={Ron Q. Nguyen and Peiyu Qin and Hai-Tian Wu and Sparsh Mishra and Tobias Wolf and Joseph Roll and Erin Morissette and Naiyuan J. Zhang and Sarah Alkidim and Kenji Watanabe and Takashi Taniguchi and Aaron W. Hui and Dima E. Feldman and Allan MacDonald and J. I. A. Li},
      year={2026},
      eprint={2607.05039},
      archivePrefix={arXiv},
      primaryClass={cond-mat.mes-hall},
      url={https://arxiv.org/abs/2607.05039}, 
}

@article{Choi_2025,
   title={Superconductivity and quantized anomalous Hall effect in rhombohedral graphene},
   volume={639},
   ISSN={1476-4687},
   url={http://dx.doi.org/10.1038/s41586-025-08621-y},
   DOI={10.1038/s41586-025-08621-y},
   number={8054},
   journal={Nature},
   publisher={Springer Science and Business Media LLC},
   author={Choi, Youngjoon and Choi, Ysun and Valentini, Marco and Patterson, Caitlin L. and Holleis, Ludwig F. W. and Sheekey, Owen I. and Stoyanov, Hari and Cheng, Xiang and Taniguchi, Takashi and Watanabe, Kenji and Young, Andrea F.},
   year={2025},
   month=Mar, pages={342–347} }

@article{Aronson2025,
  title = {Displacement Field-Controlled Fractional Chern Insulators and Charge Density Waves in a Graphene/hBN Moir\'e Superlattice},
  author = {Aronson, Samuel H. and Han, Tonghang and Lu, Zhengguang and Yao, Yuxuan and Butler, Jackson P. and Watanabe, Kenji and Taniguchi, Takashi and Ju, Long and Ashoori, Raymond C.},
  journal = {Phys. Rev. X},
  volume = {15},
  issue = {3},
  pages = {031026},
  numpages = {9},
  year = {2025},
  month = {Jul},
  publisher = {American Physical Society},
  doi = {10.1103/75gl-jzl6},
  url = {https://link.aps.org/doi/10.1103/75gl-jzl6}
}

@misc{hua2026multiknobswitchablechiralsuperconductivity,
      title={Multi-Knob Switchable Chiral Superconductivity Quartet in Rhombohedral Graphene}, 
      author={Zhenqi Hua and Shenyong Ye and Phatthanon Pattanakanvijit and Gang Shi and Tonghang Han and Emily Aitken and Jixiang Yang and Junseok Seo and Haoyang Liu and Ran Hao and Kaitai Xiao and Jiaxing Guo and Vo Tien Phong and Kenji Watanabe and Takashi Taniguchi and Chunli Huang and Cyprian Lewandowski and Long Ju and Peng Xiong and Zhengguang Lu},
      year={2026},
      eprint={2607.06520},
      archivePrefix={arXiv},
      primaryClass={cond-mat.supr-con},
      url={https://arxiv.org/abs/2607.06520}, 
}

@misc{geier2024chiraltopologicalsuperconductivityisospin,
      title={Chiral and topological superconductivity in isospin polarized multilayer graphene}, 
      author={Max Geier and Margarita Davydova and Liang Fu},
      year={2024},
      eprint={2409.13829},
      archivePrefix={arXiv},
      primaryClass={cond-mat.supr-con},
      url={https://arxiv.org/abs/2409.13829}, 
}

% ------------------------------------------------------------------------ %
%  SUPPLEMENTAL MATERIAL %
% ------------------------------------------------------------------------ %

\onecolumngrid
\newpage
\makeatletter 

\begin{center}
\textbf{\large Supplemental materials for: \\``\@title ''} \\[10pt]
Ahmed Abouelkomsan$^{1*}$, Filippo Gaggioli$^{1*}$, Daniele Guerci$^{1*}$ and Liang Fu$^1$ \\
\textit{$^1$Department of Physics, Massachusetts Institute of Technology, Cambridge, MA-02139, USA}\\
\end{center}
$^*$These authors contributed equally to this work.
\vspace{10pt}

\setcounter{page}{1} % Start from page 1 (you can use 0, but journals start at 1)
\setcounter{figure}{0}
\setcounter{section}{0}
\setcounter{equation}{0}

\renewcommand{\thefigure}{S\@arabic\c@figure}
\makeatother

\appendix

\section{Microscopic Model and Correlation Effects in Rhombohedral Multilayer Graphene}

For rhombohedral \(L\)-layer graphene, we consider the single-particle Hamiltonian:
\begin{equation}\label{sm:Llayer}
    h_L =
    \begin{pmatrix}
        h^{(0)}_1 & h^{(1)} & h^{(2)} & 0 & \cdots & 0 \\
        h^{(1)\dagger} & h^{(0)}_2 & h^{(1)} & h^{(2)} & \ddots & \vdots \\
        h^{(2)\dagger} & h^{(1)\dagger} & h^{(0)}_3 & h^{(1)} & \ddots & 0 \\
        0 & h^{(2)\dagger} & h^{(1)\dagger} & h^{(0)}_4 & \ddots & h^{(2)} \\
        \vdots & \ddots & \ddots & \ddots & \ddots & h^{(1)} \\
        0 & \cdots & 0 & h^{(2)\dagger} & h^{(1)\dagger} & h^{(0)}_L
    \end{pmatrix},
\end{equation}
where, for simplicity, we dropped the momentum dependency, $u_\ell =D (\ell-(N+1)/2)$ with $D$ displacement field and: 
\begin{equation}
    h^{(0)}_l(\k) = \begin{pmatrix}
        u_l & -t_0 f_{\k} \\ 
        -t_0 f^*_{\k} & u_l 
    \end{pmatrix},\quad
    h^{(1)}(\k) = \begin{pmatrix}
        t_4 f_{\k} & t_3 f^*_{\k} \\ 
        t_1  & t_4 f_{\k} 
    \end{pmatrix},\quad    h^{(2)} = \begin{pmatrix}
        0 & t_2/2  \\ 
        0 & 0 
    \end{pmatrix},\quad
    f_{\k}=\sum^3_{j=1}e^{i\bm u_j\cdot\k}.
\end{equation}
In the latter expression, we have introduced the vectors connecting the two sublattices $\bm u_j=a_GR^{j-1}_{2\pi/3}(0,1)/\sqrt{3}$, $R_{2\pi/3}$ rotation of $2\pi/3$ around $z$ and $a_G=0.246\,$nm.
The model parameters are taken from Ref.~\cite{Dong_2024}: \(t_0=3.1\,{\rm eV}\), \(t_\perp=t_1=380\,{\rm meV}\), and \((t_2,t_3,t_4)=(-21,290,141)\,{\rm meV}\), while the layer-dependent onsite potential \(u_\ell\) accounts for the applied displacement field.
The model features a characteristic length scale \(\ell=\hbar v_F/t_1=a_G\sqrt{3}t_0/(2t_1) \approx 1.738\,{\rm nm}\), which we use as the unit of length throughout our calculations.

\subsection{A Minimal Model for Interacting Phases in Rhombohedral Graphene}

We investigate interaction-driven phases in rhombohedral graphene within a minimal model of spin- and valley-polarized electrons. We focus on the interplay between the band dispersion and strong electronic interactions, setting aside band topological effects to reduce the complexity of the problem. 
Within our model, the electrons reside in a single band with dispersion relation $\epsilon(\k)$ and interact through a two-body potential $V(\r)$. 
In first quantization, the $N$-particle Hamiltonian reads
\begin{equation}\label{Hamiltonian}
    H = \sum_{j=1}^{N} \epsilon(-i\hat \nabla_j)+\sum_{i<j}V(\r_i-\r_j),
\end{equation}
where $V(\r)$ is the Coulomb interaction:
\begin{equation}
    V(\r)=\frac{e^2}{4\pi\epsilon\epsilon_0\ell }\frac{\ell}{r}=\frac{\ell E_C}{r},\quad V_\q=\frac{2\pi\ell^2E_C}{q},\quad \q = \ell\mathbf q,
\end{equation}
where lengths are measured in units of $\ell$, and wave vectors in units of $\ell^{-1}$. In the following, we measure the energy in units of $E_C=e^2/(4\pi\epsilon_0\epsilon\ell)$, where $e^2/(4\pi\epsilon_0\ell)\approx0.8285$eV.

In Eq.~\eqref{Hamiltonian}, we have also considered the dispersion relation $\epsilon(\k)$. 
In our Hartree-Fock (HF) calculations, this is taken as the dispersion relation of the first band above charge neutrality of tetralayer graphene~\eqref{sm:Llayer}. 

\begin{figure}
    \centering
    \includegraphics[width=\linewidth]{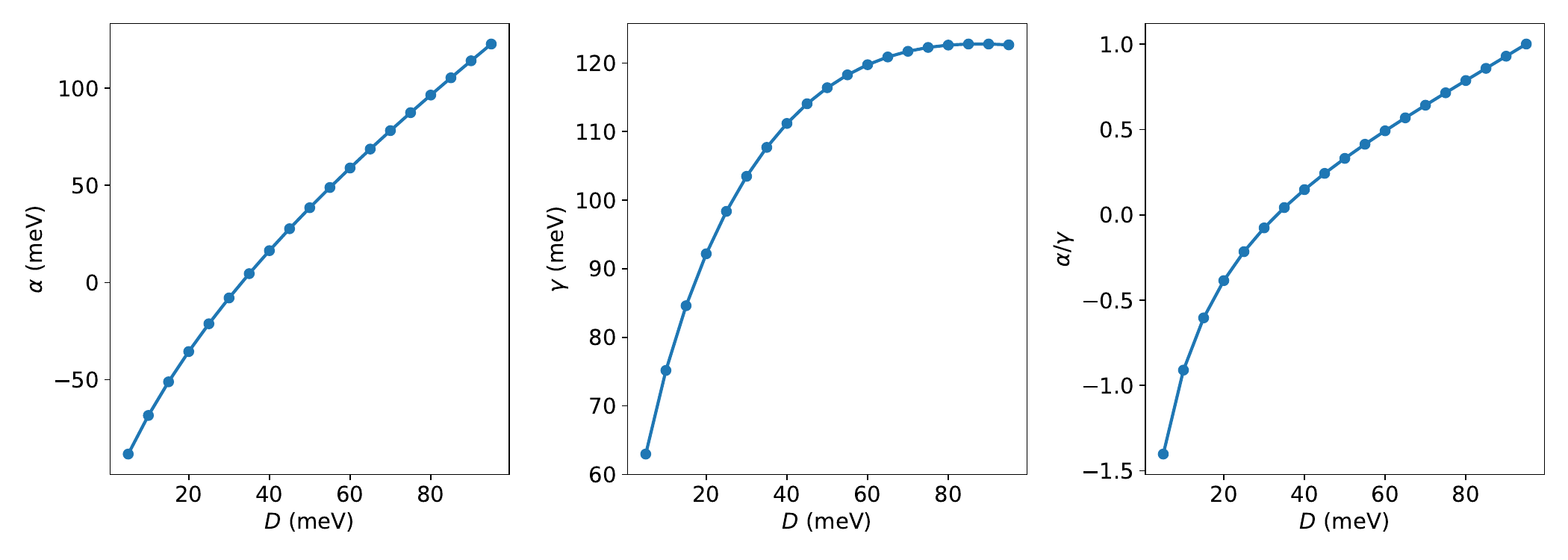}
    \caption{Fitted values of the parameters $\alpha$ and $\gamma$ extracted from the first conduction band above charge neutrality of R4G.}
    \label{fig:kpfitting}
\end{figure}

\begin{figure}
    \centering
    \includegraphics[width=.7\linewidth]{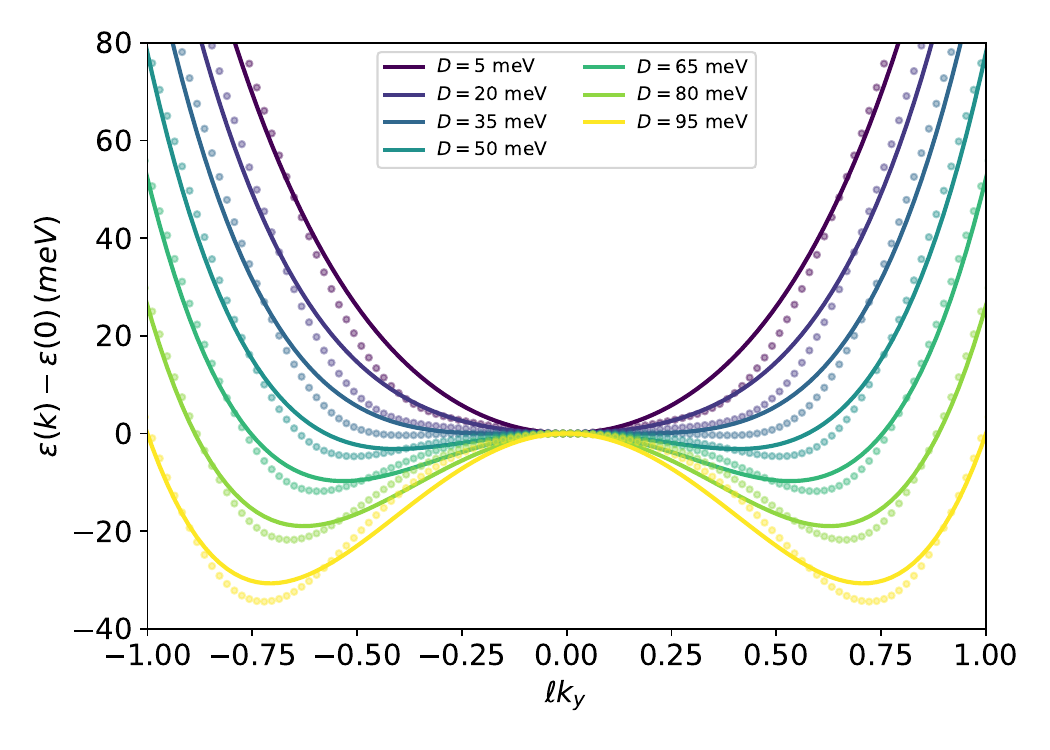}
    \caption{Four layers rhombohedral graphene band structure for different displacement fields. Dots show the bandstructure of the full model while solid lines the fit.}
    \label{fig:bands}
\end{figure}

On the other hand, for the NN-VMC and additional HF simulations we considered a ``Mexican hat" dispersion relation.
To reproduce the properties of rhombohedral graphene, the coefficients of the de also included the fit of the bandstructure of tetralayer graphene, we express $\alpha$ and $\gamma$ in the same units of energy as: 
\begin{equation}\label{eps_k}
    \epsilon({\k})=-\bar\alpha\left(\frac{ a_G}{\ell}\right)^2(\ell\mathbf k)^2+\bar\gamma \left(\frac{a_G}{\ell}\right)^4(\ell\mathbf k)^4,\quad \k=\ell\mathbf k,\quad 
\end{equation}
and, as a result, we have: 
\begin{equation}
\label{eq:alphaandgamma}
    \alpha=\bar\alpha\left(\frac{ a_G}{\ell}\right)^2,\quad \gamma=\bar\gamma \left(\frac{ a_G}{\ell}\right)^4.
\end{equation}
Finally, for each value of the displacement field, we determine $\alpha$ and $\gamma$ through a least-squares fit. The resulting fits are shown in Fig.~\ref{fig:kpfitting} and in Fig.~\ref{fig:bands}.

\subsection{Interaction strength and correlation regime}

\begin{figure}
    \centering
    \includegraphics[width=0.8\linewidth]{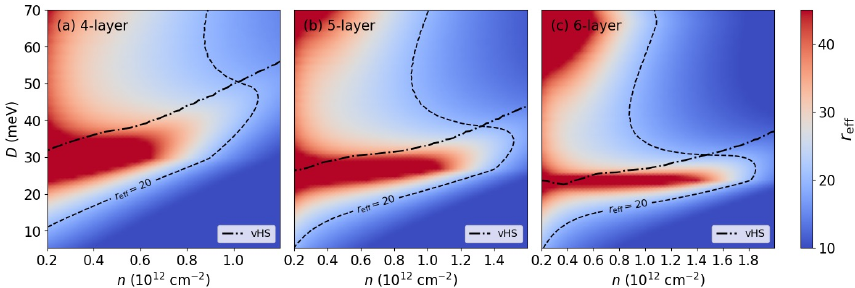}
    \caption{The effective interaction parameter $r_{\rm eff}$ for 4, 5 and 6 layer rhombohedral graphene, with relative dielectric constant $\epsilon=5$. We show the van Hove singularity line.}
    \label{fig:RL6Layers}
\end{figure}

We introduce a simple guide to quantify the degree of correlation by comparing the average kinetic energy and the average Coulomb energy per particle:
\begin{equation}\label{non_interacting_average}
    E_{\rm kin}
    =
    \frac{1}{ \ell^2 n}
    \int \frac{d^2\k}{(2\pi)^2}
    f(\k)\epsilon(\k),
    \qquad
    E_{\rm int}
    =
    \frac{E_C\ell}{d_e},
    \qquad
     \ell^2 n
    =
    \int \frac{d^2\k}{(2\pi)^2}f(\k),\quad f(\k)=\frac{1}{e^{\frac{\epsilon(\k)-\mu}{k_BT}}+1},
\end{equation}
where $\k$ is dimensionless and the dispersion of the conduction of band of the Hamiltonian~\eqref{sm:Llayer} is measured relative to the band minimum, so that $E_{\rm kin}\ge0$.
The last equation fixes the chemical potential, and  $ d_{e}=\sqrt{1/(\pi  n)}$ is the interparticle distance. 
We define the dimensionless ratio $r_{\rm eff}={E_{\rm int}/E_{\rm kin}}$ to qualitatively quantify the degree of electronic correlations. 
The latter quantity together with the line of the van Hove singularity is shown in Fig.~\ref{fig:RL6Layers}.

\begin{figure}
    \centering
    \includegraphics[width=.6\linewidth]{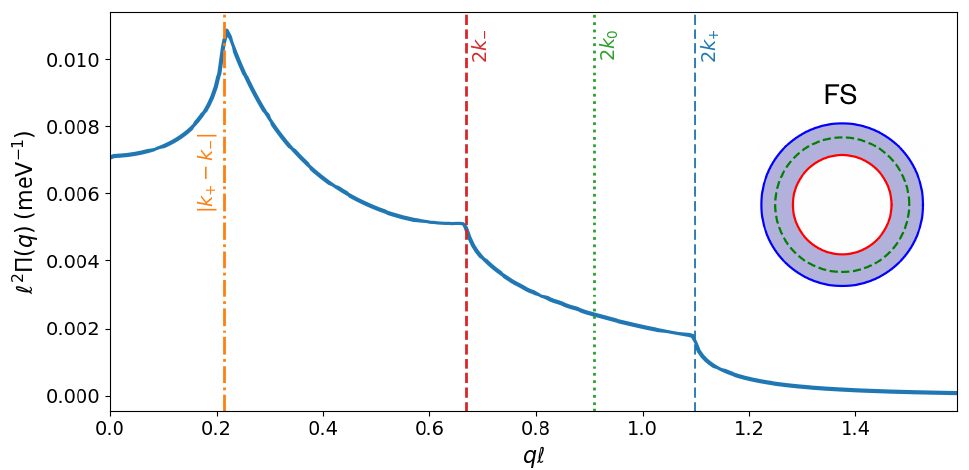}
    \caption{Static density-density susceptibility as a function of $q$ and corresponding annular Fermi surface for $D=55$meV and $n=0.5\times 10^{12}$cm$^{-2}$. 
        Vertical lines indicate the characteristic momentum scales.}
    \label{fig:Piq}
\end{figure}

\subsection{Non-interacting density-density susceptibility}

To characterize the intrinsic length scales of the annular Fermi gas with Mexican-hat dispersion, we compute the static density–density susceptibility:
\begin{equation}
    \Pi(\q)=\frac{1}{\ell^2}\int\frac{d^2\k}{(2\pi)^2}\frac{f(\epsilon({\k+\q})-\mu)-f(\epsilon(\k)-\mu)}{\epsilon({\k})-\epsilon({\k+\q})},
\end{equation}
where $\k$ is expressed in unit of $1/\ell$ and we have introduced the chemical potential $\mu$ fixing the density of electrons $n$ (see Eq.~\eqref{non_interacting_average}). 
Figure~\ref{fig:Piq} shows the density--density susceptibility as a function of momentum $q$ for the Mexican-hat dispersion defined in Eq.~\eqref{eps_k} at $D=55\,\mathrm{meV}$ and $n=0.5\times10^{12}\,\mathrm{cm}^{-2}$. 
The vertical blue and red lines indicate the characteristic momenta $2k_+$ and $2k_-$ associated with the outer and inner Fermi surfaces, respectively, while the green and orange lines mark $2k_0$ and $\sqrt{k_+^2+k_-^2-2k_+k_-}=|k_+-k_-|$. 

We emphasize that the charge ordering wave vector of the self-doped Phantom crystals we observe does not correspond to any of the characteristic momenta shown in Fig. ~\ref{fig:Piq} reflecting that this state does not occur in the weakly interacting reime.

\section{Neural Network Variational Monte Carlo}
\begin{figure}
    \centering
    \includegraphics[width=0.5\linewidth]{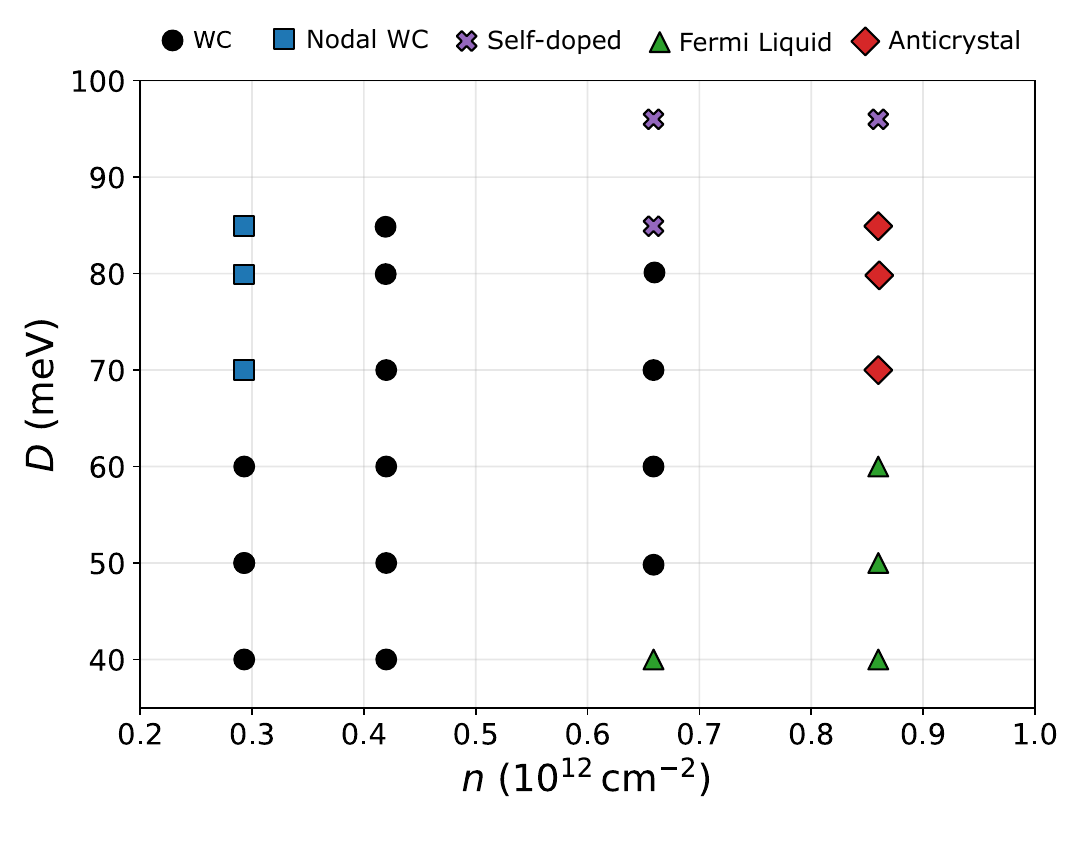}
    \caption{NN-VMC phase diagram of tetralayer graphene modeled using the Hamiltonian \eqref{eq:alphaandgamma}. Calculations are done using relative dielectric constant $\epsilon = 5$ with $N = 25$ particles. We use a triangular supercell with equal aspect ratio.}
\label{fig:NNphasediagram}
\end{figure}

We perform neural network variational Monte Carlo simulations on the Hamiltonian~\eqref{Hamiltonian} with the dispersion relation in Eq.~\eqref{eps_k}.
 We use self-attention neural network following Ref. \cite{Geier_2025}. In Table \ref{Tab:Hyperparams}, we list the hyperparameters used in our simulations. Fig. \ref{fig:NNphasediagram} shows the phase diagram obtained from NN-VMC for the tetralayer case as a function of the density $n$ and the inferred displacement field $D$ obtained through fitting $\alpha$ and $\gamma$ in equation \eqref{eq:alphaandgamma} to the full dispersion.

\begin{figure}
    \centering
    \includegraphics[width=0.9\linewidth]{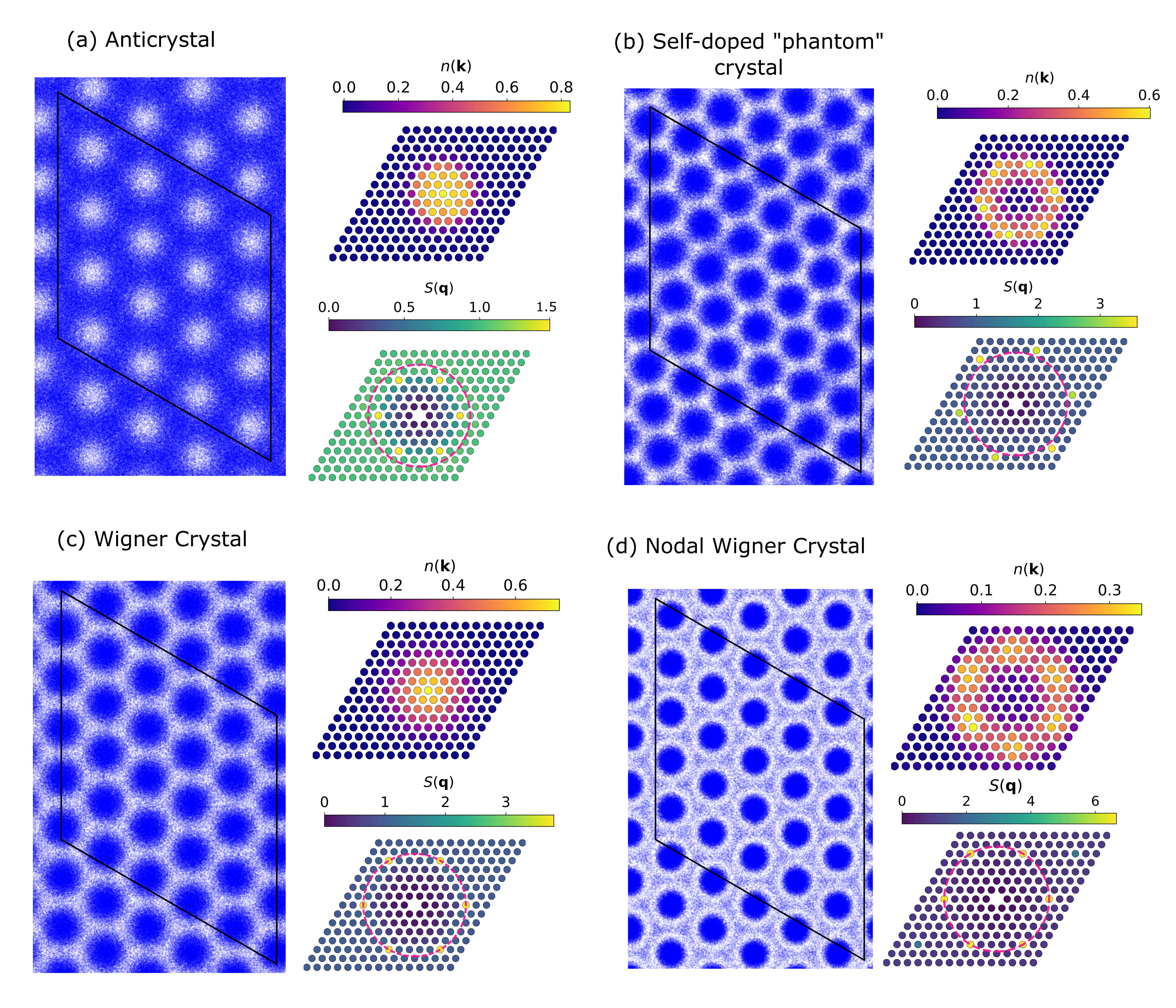}
    \caption{Crystalline states found in NN-VMC calculations of the Hamiltonian \eqref{Hamiltonian} with the dispersion relation \eqref{eq:alphaandgamma}. Each panel shows the density (left) where blue denotes regions of high electron density and white denote region of low density in addition to the momentum occupation $n(\k)$ structure factor $S(\q)$. The black parallelogram represents the simulation supercell. Calculations are done for $N = 25$ particles with relative dielectric constant $\epsilon = 5$. The red circle in the structure factor corresponds the momentum $|\K_{\rm WC}|$ corresponding to a WC crystal unit cell with one electron. 
    The parameters are $(\alpha/E_C, \gamma/E_C, n)=(0.582,0.740,0.86\times10^{12}~{\rm cm}^{-2})$ for (a), $(0.740,0.740,0.659\times10^{12}~{\rm cm}^{-2})$ for (b), $(0.232,0.702,0.293\times10^{12}~{\rm cm}^{-2})$ for (c), and $(0.544,0.739,0.293\times10^{12}~{\rm cm}^{-2})$ for (d). }
\label{fig:NN_observables}
\end{figure}

Our calculations reveal a variety of different crystalline phases.  In Fig. \ref{fig:NN_observables}, we show representative charge densities of such states in addition to the momentum occupation function $n(\k)$ and the structure factor $S(\q)$. 
 
 \begin{table}
    \centering
    \renewcommand{\arraystretch}{1.3}
    \begin{tabular}{ll l}
        \toprule
        \textbf{Parameter} & & \textbf{Value} \\
        \midrule
        \textbf{Architecture} & & \\ 
        Network layers & & $L = 4$ \\
        Attention heads per layer & & $N_{\rm heads} = 4$ \\
        Attention dimension & & $d_{\rm attn} = d_{\rm attnvals} = 32$ \\
        Perceptron dimension & & $d_L = 128$ \\
        $\#$ perceptrons per layer & & $2$ \\
         Determinants & & $N_{\rm det} = 2$\\
         Layer norm & & True \\
         Activation & & GELU \\
        \midrule
        \textbf{Training} & & \\ 
        Training iterations & & $60{\rm k} - 150 {\rm k}$ \\
        Learning rate (fixed) &  & $\eta_0 = 0.1$  \\
        Optimizer &  & KFAC \\
        \midrule
        \textbf{MCMC} & & \\
        Batch size & & $1024-2048$ \\
        \midrule
        \textbf{KFAC} & & \\ 
        Norm constraint & & $1 \times 10^{-3}$ \\
        Damping & & $1 \times 10^{-3}$ \\
        $L_2$ regularization & & 0.0 \\
        Momentum & & 0.0 \\
        \bottomrule
    \end{tabular}
    \caption{Table of default hyperparameters used in our numerical calculations.}
    \label{Tab:Hyperparams}
\end{table}

\section{Hartree-Fock Solution}

At Hartree-Fock level, the expectation value of the interacting Hamiltonian, presented in the main text, is expressed in terms of the one-particle reduced density matrix (1RDM) $n_{\k,\k'}$ only:
\begin{equation}\begin{split}\label{Hartree-Fock}
    E[\hat n] =\mel{\Psi_{\rm HF}}{H}{\Psi_{\rm HF}}=\sum_\k\epsilon(\k) n_{\k,\k} + \frac{1}{2A}\sum_{\k\k'\q\neq0}V_{\q}\left[ n_{\k,\k+\q}n_{\k',\k'-\q}-n_{\k,\k'-\q}n_{\k',\k+\q}\right],
\end{split}\end{equation}
where $V_\q=2\pi \ell^2 E_C/q$ is the Fourier transform of the Coulomb interaction, $A$ the area of the two-dimensional system and $n_{\k,\k'}=\mel{\Psi_{\rm HF}}{c^\dagger_{\k}c_{\k'}}{\Psi_{\rm HF}}$ with $c_{\k}$ annihilation operator of an electron with momentum $\k$ and energy $\epsilon(\k)$. In the following, we measure momenta in units of $1/\ell$ and lengths in units of $\ell$.  

Minimizing this energy functional~\eqref{Hartree-Fock} with respect to $n_{\k,\k'}$ yields the optimal Slater determinant $\ket{\Psi_{\rm HF}}$ approximating the many-body ground state of the Hamiltonian.
In the following, we consider three Hartree-Fock schemes: restricted calculations for liquid phases, restricted calculations for phases with triangular unit cell, and fully unconstrained Hartree-Fock calculations. The combination of these approaches comparing the energies of different Hartree-Fock minima allows us to establish the Hartree-Fock phase diagram.

\subsection{Translationally invariant liquids}

For a translationally invariant liquid, $n_{\k,\k'}$ is diagonal, and the Hartree-Fock energy per particle reduces to: 
\begin{equation}
    E[\hat n]=\sum_{\k}\epsilon(\k) n_{\k}+\frac{1}{2A}\sum_{\k\p}^{\k\neq\p}n_\k n_\p\left[V_0-V_{\k-\p}\right],
\end{equation}
where $n_{\k}$ is the $\k$ space occupation number.  
The Hartree contribution is independent of the momentum distribution,
\begin{equation}
    E_H[\hat n]
    =
    \frac{N_e(N_e-1)}{2A}V_0 ,
\end{equation}
and is therefore removed as an irrelevant constant energy shift. 
In the thermodynamic limit the Hartree-Fock energy per particle becomes: 
\begin{equation}
    \epsilon[\hat n]=\frac{1}{\ell^2 n}
    \int \frac{d^2\k}{(2\pi)^2}
    n_{\k}\epsilon(\mathbf k)-\frac{\pi E_C}{\ell^2 n}\int \frac{d^2\k d^2\p}{(4\pi^2)^2}\frac{n_{\k}n_{\p}}{|\k-\p|},
\end{equation}
Minimizing this functional with respect to the occupation function $n_{\mathbf k}$, subject to the density constraint, determines the optimal translationally invariant Hartree-Fock state.
The minimization yields metallic states that can be either isotropic or nematic.
Nematic solutions develop in the annular Fermi surface regime.

\subsection{Restricted Hartree-Fock}

Restricted calculations are performed assuming the system develops a periodicity with lattice vectors $\a_1$, $\a_2$ and unit cell area $A_{\rm UC}=\a_1\times\a_2$. 
Moreover, we consider a triangular unit cell defined by primitive vectors $\a_1$ and $\a_2$ of equal length, $|\a_1|=|\a_2|$, and relative angle between $\a_2$ and $\a_1$ of $2\pi/3$.
Correspondingly, we have the reciprocal lattice vectors 
\begin{equation}
    \G_1=-2\pi\frac{\mathbf z\times \a_2}{\a_1\times\a_2},\quad \G_2=2\pi\frac{\mathbf z\times \a_1}{\a_1\times\a_2}.
\end{equation}
Under these assumptions, the 1RDM takes the form $n([\k])_{\G,\G'}$ where $[\k]\in\rm BZ$, $\G$ and $\G'$ are reciprocal lattice vectors.
Moreover, the single particle wavefunction takes the form: 
\begin{equation}
    \psi_{\k n}(\r)=e^{i\k\cdot\r}\sum_{\G}e^{i\G\cdot\r}U_{\G,n}(\k).
\end{equation}
In terms of the Hartree-Fock basis $U_{\G,n}([\k])$, we define the one particle reduced density matrix: 
\begin{equation}\label{1RDM}
\left[n([\k])\right]_{\G,\G'}=\langle c^\dagger_{[\k],\G}c_{[\k],\G'}\rangle=\sum_{n=1}^{N_b}U_{\G',n}([\k])f(E_{n}(\k)-\mu)U^\dagger_{n,\G}([\k]),
\end{equation}
where $c_{[\k],\G}$ destroys a plane wave at $\k=[\k]+\G$, $N_b$ is the number of bands, $E_{n}(\k)$ is the Hartree-Fock dispersion and $\mu$ is the chemical potential.

The resulting Hartree-Fock energy per particle becomes: 
\begin{equation}\begin{split}
    \epsilon[\hat n] =&\frac{1}{\ell^2 n}\int\frac{d^2\k}{(2\pi)^2}\sum_{\G}\epsilon([\k]+\G)n([\k])_{\G,\G}+\frac{1}{2\ell^2 n}\sum_{\G\neq 0}V_{\G}\rho(\G)\rho(-\G)\\
    &-\frac{A^2_{\rm UC}}{2\ell^2 n}\int_{\rm BZ}\frac{d^2\k }{(2\pi)^2}\frac{d^2\k'}{(2\pi)^2}\sum_{\G\G'\Q}V_{\k-\k'-\Q}n([\k])_{\G,\G'}n([\k'])_{\G'+\Q,\G+\Q},
\end{split}\end{equation}
where we have introduced the Fourier components of the charge density: 
\begin{equation}
    \rho(\G)=A_{\rm UC}\int\frac{d^2\k}{(2\pi)^2}\sum_{\G'}n([\k])_{\G',\G'+\G}.
\end{equation}
Finally, for a fixed electron density $n$, we compute Hartree--Fock solutions as a function of the unit-cell area $A_{\rm UC}$ and orientation, and determine the optimal state by minimizing the energy with respect to $A_{\rm UC}$.

For fixed density, displacement field, and relative dielectric constant $\epsilon$, we determine the minimum-energy solution by solving the Hartree-Fock equations at fixed unit-cell area $A_{\rm UC}$, followed by a minimization of the energy with respect to \(A_{\rm UC}\).

\subsection{Unrestricted Hartree-Fock}

Additionally, to explore the full range of possible translational-symmetry-broken patterns, we perform unconstrained Hartree–Fock simulations.
To minimize the Hartree-Fock energy, we compute the energy variation with respect to $n_{\k,\k'}$: 
\begin{equation}
    \Delta E[\{\hat n\}]=\sum_{\k,\k'}\frac{\delta E[\{\hat n\}]}{\delta n_{\k,\k'}}\delta n_{\k,\k'},
\end{equation}
where we have introduced:
\begin{equation}\begin{split}
   F_{\k,\k'}&= \frac{\delta E[\{\hat n\}]}{\delta n_{\k,\k'}}=\delta_{\k,\k'}\epsilon(\k) + \frac{1}{A}\sum_{\p \p'} n_{\p,\p'}\left[V_{\k\p,\p'\k'}-V_{\k\p,\k'\p'}\right],
\end{split}\end{equation}
and we have introduced $V_{\k_1\k_2,\k_3\k_4} =V_{\k_1-\k_4}\delta_{\k_1+\k_2-\k_3-\k_4,0}$. 

The matrix $n_{\k,\k'}$ defines the set of variational parameters. 
First, we note that $\k$ and $\k'$ are unconstrained discrete variables, each taking values in an infinite set.
We introduce the UV cutoff $\Lambda$, keeping only momenta $\k$ with $k<\Lambda$. 
Moreover, $\hat n$ satisfies the following conditions: 
\begin{equation}
    n_{\k,\k}=\mel{\Psi}{c^\dagger_{\k} c_{\k}}{\Psi}=\|c_{\k}\ket{\Psi}\|^2\ge 0,\quad \Tr \hat n = N_e. 
\end{equation}

To this end, we introduce the rectangular matrix $C_{\k,j}$, where
$\k$ runs over $N_\Lambda$ momenta and $j=1,\dots,N_e$
labels the occupied single-particle orbitals. Assuming that the
$j$-basis consists of non-orthogonal orbitals, we find:
\begin{equation}
    [\hat n]^T= C\cdot S^{-1}\cdot C^\dagger, 
\end{equation}
where we have introduced the matrix $S=C^\dagger \cdot C$. 
Notice that: 
\begin{equation}
    \Tr \hat n=\sum_{\k}\sum_{jj'=1}^{N_e} C_{\k,j}S^{-1}_{jj'} C^\dagger_{j',\k}=\sum_{j=1}^{N_e}\left(S^{-1}\cdot S\right)_{jj}=N_e,\quad \forall\, C_{\k,j}. 
\end{equation}

Within this formulation, the derivative of the energy with respect to $C^*_{\k,j}$ reads: 
\begin{equation}\label{hartree-fock}\begin{split}
    \frac{\delta E}{\delta C_{\p,j}^*}&=\sum_{\k\k'}(\mathbf 1-C\cdot S^{-1}\cdot C^\dagger)_{\p\k} F_{\k,\k'}C_{\k',n}S^{-1}_{nj},\\
    E[\hat n] &= \frac{1}{2}\sum_{\k,\k'}\left[\delta_{\k,\k'}\epsilon(\k) + F_{\k,\k'}\right] n_{\k,\k'}.
\end{split}\end{equation}
Hence, given $C_{\k,j}$, we can compute both the gradient and the average value of the energy, which can then be utilized for off-the-shelf minimization. 
At the Hartree-Fock minimum, we have: 
\begin{equation}
    [F,P]=0,\quad P=n^T.
\end{equation}

\subsection{Summary of Hartree-Fock solutions}

In the following, we illustrate the energy landscape for different crystalline ground states obtained within restricted and unrestricted Hartree-Fock, and their respective band structure. 
The Hartree–Fock simulations presented below employ the dispersion relation $\epsilon(\k)$ of the lowest conduction band above charge neutrality in rhombohedral tetralayer graphene.

\begin{figure}
    \centering
    \includegraphics[width=0.5\linewidth]{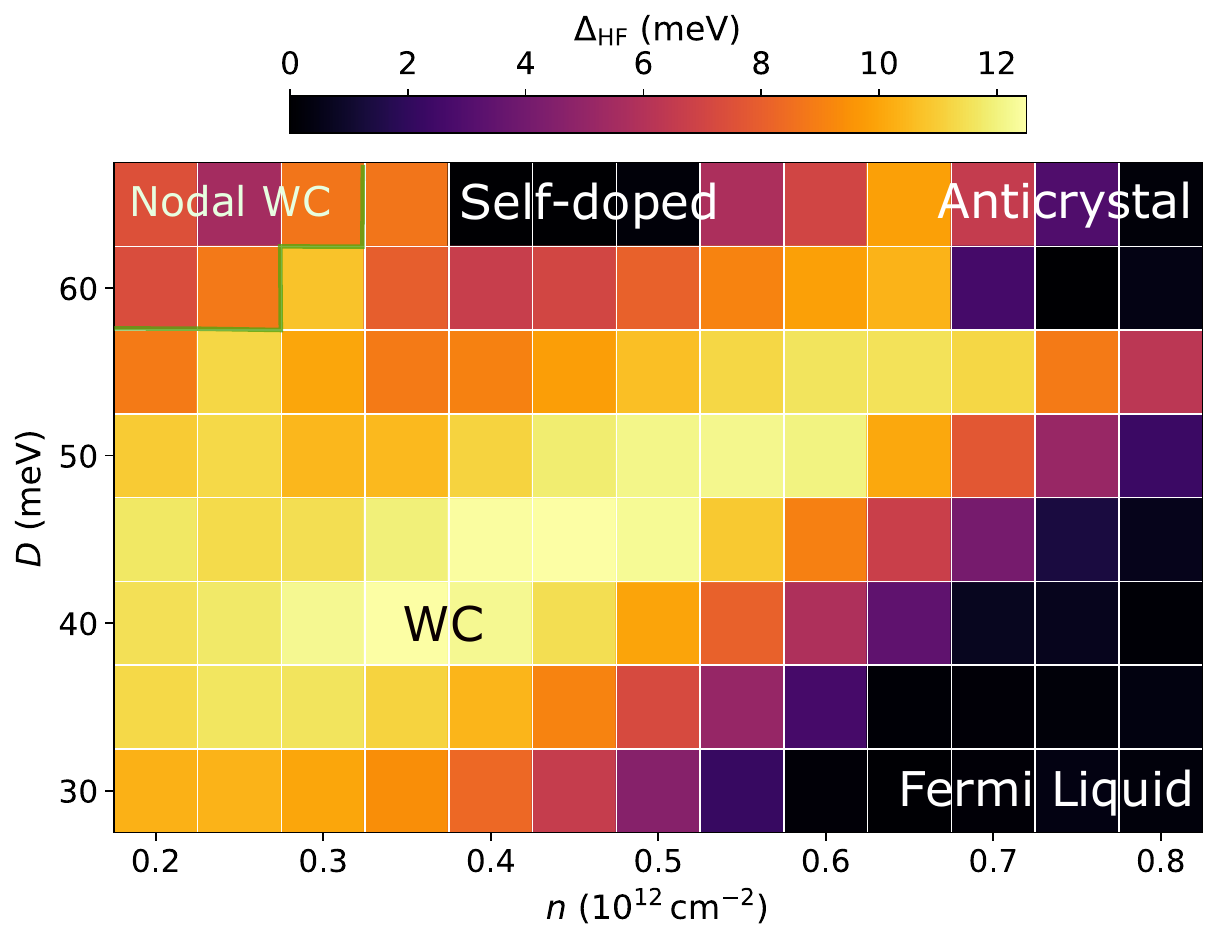}
    \caption{HF charge gap $\Delta_{\rm HF}$ (Eq. \eqref{eq:HFchargegap} as a function of displacement field $D$ and density $n$ for tetralayer rhombohedral graphene. }
    \label{fig:HFchargegap}
\end{figure}
In addition to Fig. 3 in the main text which plots the filling factor $\nu$ of the HF ground state, we present in Fig. \ref{fig:HFchargegap} the Hartree--Fock charge gap, defined as

\begin{equation}
\label{eq:HFchargegap}
    \Delta_{\rm HF}=\min_{\k}[E_{\rm unocc}(\k)] -\max_{\k}[E_{\rm occ}(\k)]
\end{equation}

The HF charge gap allows us to identify the metallic and insulating states we find which we analyze below.

\subsubsection{Wigner crystal}

In this section, we focus on translational-symmetry-broken insulating states, namely Wigner crystals. 
We compare the conventional Wigner crystal (WC), realized at small displacement field, with the nodal Wigner crystal (nWC), observed at larger displacement fields, emphasizing their distinct real-space charge-density profiles.

\begin{figure}
    \centering
    \includegraphics[width=\linewidth]{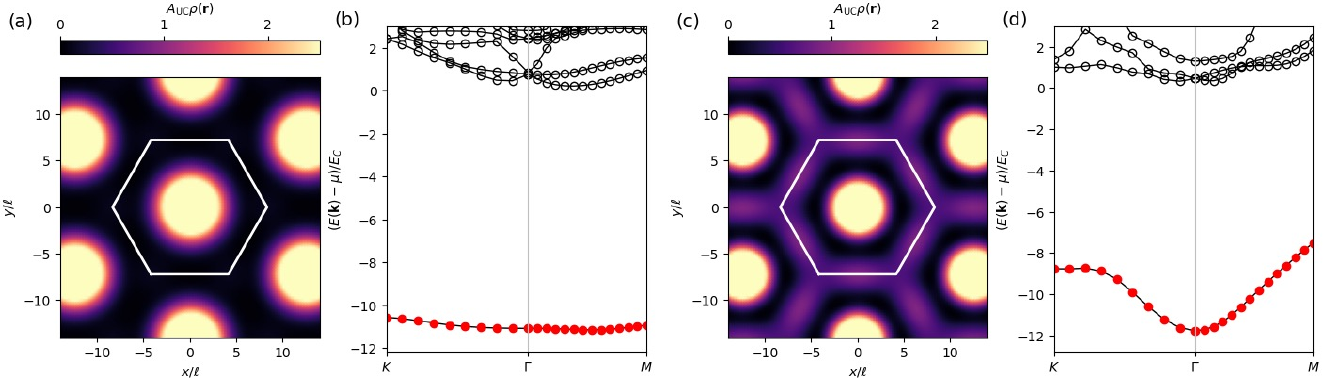}
    \caption{Hartree-Fock charge density (a,c) and band structure (b,d) of the Wigner crystal (left panels) and nodal Wigner crystal (right panels). Calculations are performed for $n=0.18\times10^{12}$cm$^{-2}$ and displacement fields $D=45$meV for the Wigner crystal and $D=60$meV for the nodal Wigner crystal.}
    \label{fig:WC_nWC}
\end{figure}

Fig.~\ref{fig:WC_nWC}(a) and Fig.~\ref{fig:WC_nWC}(b) show the Hartree-Fock solution in the Wigner crystal (WC) regime.  
The single-particle spectrum features an isolated flat band separated by a gap from the remaining bands of the order of $10$meV, while the charge density forms a triangular-lattice pattern with the emergent unit cell indicated in white in Fig.~\ref{fig:WC_nWC}(a).The restricted Hartree-Fock calculations are confirmed by unrestricted results obtained using the same values of density and displacement field.

\subsubsection{Nodal Wigner crystal}

Increasing the displacement field in the low-density annular Fermi-surface regime stabilizes a distinct Wigner crystal state that we call a nodal Wigner crystal (nWC). 
As an illustration, in Fig.~\ref{fig:WC_nWC}(c) and Fig.~\ref{fig:WC_nWC}(d), we show results for $\rho=0.18\times10^{12}\,\mathrm{cm}^{-2}$, although this regime extends over a finite range of densities and displacement fields. 
At the Hartree-Fock level, the state shows a single particle spectrum with a gap of the order of $8$meV. 

Fig.~\ref{fig:WC_nWC}(a) and Fig.~\ref{fig:WC_nWC}(c) show distinct spatial patterns in the WC and nWC obtained for small and large values of the displacement field, respectively.
The distinctive feature of the nWC is its nonmonotonic real-space charge-density profile: The density first develops a hexagonal ring of local minima around each charge maximum, followed by the emergence of secondary maxima between neighboring unit cells.
Within the rotational symmetric approximation, in the Mexican hat model, this minima form a ring with radius which can be obtained analytically from the first nodal ring of $J_0(k_0 r)$, $r_n =j_{0,1}/k_0\simeq 2.4048/k_0$, as detailed in the main text.

\subsubsection{Self-doped crystal}

In addition to insulating Wigner crystals, the HF calculations find a metallic crystalline phase in which the carrier density is incommensurate with the spontaneously formed lattice, resulting in a non-integer number of electrons per crystal unit cell. 
As a result, this phase is metallic and coexists with a Wigner-crystal charge-density wave.
This state is found in the strong Mexican hat regime at large displacement fields and intermediate density values. In real space, its density distribution is characterized by a number of density peaks that is different from the number of particles in the system and, hence, results in an incommensurate filling factor $\nu$.

\begin{figure}
    \centering
    \includegraphics[width=.85\linewidth]{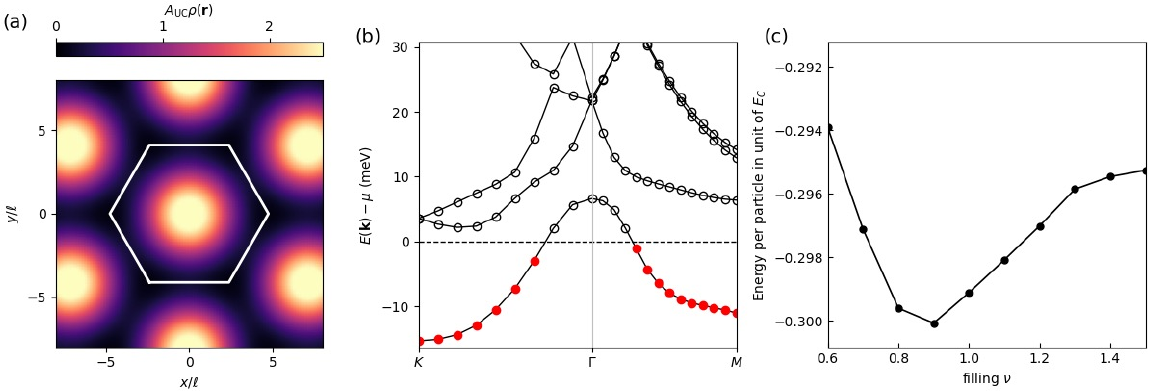}
    \caption{Hartree-Fock charge density (a), band structure (b) and energy of the self-doped Wigner crystal. Calculations are performed for $n=0.51\times10^{12}$cm$^{-2}$ and displacement fields $D=65$meV.}
    \label{fig:dopedWC}
\end{figure}

The results of the HF calculations are shown in Fig.~\ref{fig:dopedWC}(a--c). Importantly, the energy minimum at non-integer filling in Fig.~\ref{fig:dopedWC}(c) demonstrates the metallic character of the HF band structure shown in Fig.~\ref{fig:dopedWC}(b) and establishes that the global energy minimum occurs at an incommensurate lattice constant. Consequently, the period of the charge-density wave shown in Fig.~\ref{fig:dopedWC}(a) is not commensurate with the electron density, yielding a non-integer value of $\rho A_{\rm UC}$.

\subsubsection{Anti-crystals}

In this section, we focus on a translational-symmetry-broken phase that emerges in the regime at higher carrier densities and displacement fields.
In the Mexican-hat model, this regime corresponds to an annular Fermi surface for which the hole density enclosed by the inner Fermi surface and the electron density are in a $1:2$ ratio. This state persists over a broader range of carrier densities and interaction strengths.

\begin{figure}
    \centering
    \includegraphics[width=.85\linewidth]{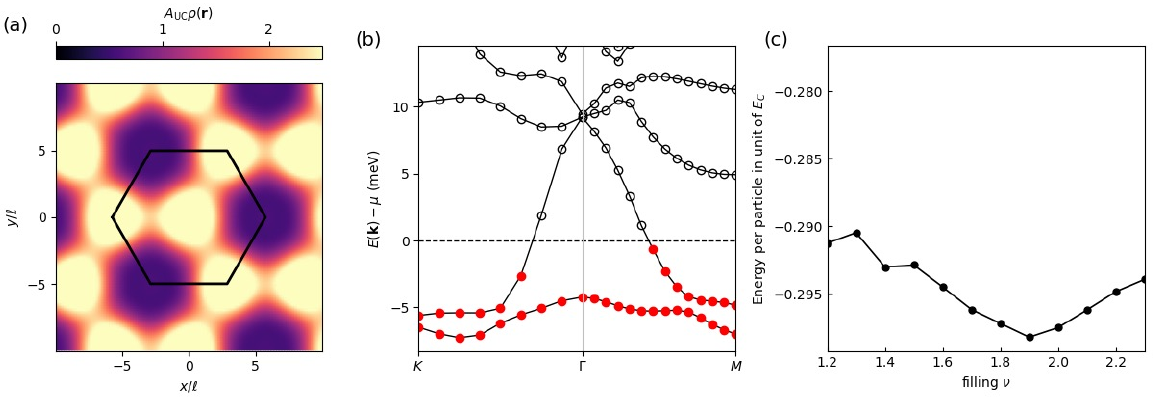}
    \caption{Hartree-Fock charge density (a), band structure (b) and energy of the anticrystal. Calculations are performed for $n=0.73\times10^{12}$cm$^{-2}$ and displacement fields $D=60$meV.}
    \label{fig:ACm}
\end{figure}

Fig.~\ref{fig:ACm} shows the Hartree-Fock solution obtained for $D=60\,{\rm meV}$ and $n=0.73\times10^{12}\,{\rm cm}^{-2}$, where the variational minimum occurs for a crystalline unit cell containing $\nu=2$ electrons. 
The resulting Hartree-Fock band structure displays two filled bands separated from the remote bands by a finite energy gap, consistent with an incompressible crystalline state.
Finally, the resulting charge-density pattern develops minima on a triangular lattice, while the maxima form a honeycomb lattice occupying the complementary Wyckoff positions.
These results are also confirmed by fully unrestricted Hartree-Fock calculations.

\section{Theoretical analysis of the nodal Wigner crystal}

At sufficiently small densities and strong repulsive interactions, the  ground state of many-body electron systems consists of a crystalline state where each particle's wave packet is tightly localized around the triangular lattice sites.
Provided that the spacing between neighboring electrons is large enough, the strong confining force exerted by the neighboring electrons can then be modeled as a simple quadratic potential, allowing us to gain much understanding of the many-body ground state from studying an effective harmonic oscillator single-particle problem around each lattice site. 
For electrons with parabolic dispersion, one finds that each particle's wave packet has Gaussian shape with characteristic width set by the Bohr radius, which shrinks to zero in the classical limit.

In this Appendix, we consider the effects of the Mexican hat dispersion on electron crystallization in the small density / strong displacement field limit, where the additional scale $k_0$ associated to the ring of dispersion minima, which we assume to be much larger than the inverse interparticle distance, will lead to interesting modifications to electronic wave packet around each crystalline lattice site \cite{joy2023wigner,joy2025chiral}.

In this limit, the effective single particle problem reads
\begin{equation}
    H = -\alpha k^2 + \gamma  k^4 +\Omega^2 r ^2,
\end{equation}
with the curvature $\Omega^2\sim E_C/\ell^2$  capturing the repulsive interaction with the surrounding charges at small electron displacements.

To proceed, we rewrite the Hamiltonian with the help of the Mexican hat momentum scale $k_0$, yielding
\begin{align}
  \frac{H}{(\gamma k_0^4)} &= \gamma k_0^4\left[-(4\gamma/\alpha) k^2 + \gamma (4\gamma/\alpha^2) k^4  +\frac{\Omega^2}{\gamma k_0^4} r ^2  \right]=\gamma k_0^4\left[-2(k^2/k_0^2) +  (k^4/k_0^4)  +\frac{\Omega^2}{\gamma k_0^4} r ^2  \right]\nonumber\\
&=2\gamma k_0^4\left[\frac{\left(k^2 - k_0^2\right)^2}{2k_0^4}+\frac{\Omega^2}{2\gamma k_0^4} r ^2  \right]
=\frac{2\gamma k_0^4}{k_0^2}\left[\frac{\left(k^2 - k_0^2\right)^2}{2k_0^2}+\frac{k_0^2\Omega^2}{2\gamma k_0^4} r ^2  \right],
\end{align}
where in going from the first to the second line we added an irrelevant constant to complete the square. 
By redefining units of energy and introducing the inverse lenght squared $\omega^2 = \Omega^2/\gamma k_0^4$, we finally obtain the simple Hamiltonian
\begin{equation}
    H = \frac{\left(k^2 - k_0^2\right)^2}{2k_0^2}+\frac{k_0^2\omega^2r ^2}{2} .
\end{equation}
We will now solve this Hamiltonian by working in momentum-basis, such that $\r = i \nabla_\k$, and using rotational symmetry to rewrite $\nabla_\k^2 = \partial_k^2 + (1/k)\partial_k + (1/k^2)\partial_\phi^2$.
This gives us (we define the effective mass $M = 1/k_0^2\omega^2$):
\begin{equation}
    H = - \frac{1}{2M} \left(\partial_k^2 + \frac{\partial_k}{k} + \frac{\partial_\phi^2}{k^2}\right) + \frac{\left(k^2 - k_0^2\right)^2}{2k_0^2}.
\end{equation}
The problem can be separated in radial and angular part, implying that $\Psi(k,\phi)=\psi(k)\chi(\phi)$. The angular part is trivially solved by $\chi_m(\phi) = e^{im\phi}$. For sufficiently large $k_0$, the radial part can instead be found by expanding $k \approx k_0 + \delta k$, with $\delta k \ll k_0$, which gives us the differential equation
\begin{equation}
    \left[-\frac{1}{2M}\left(\partial_k^2 + \frac{\partial_k}{k_0}  - \frac{m^2}{k_0^2}\right) + 2(\delta k)^2\right]\psi(\k) =E_{n,m}\,\psi(\k).
\end{equation}

Since we are interested in the ground state, we can set $m=0$ from the beginning. Indeed, the angular contribution is positive and proportional to $m^2$, so the lowest-energy state is rotationally invariant. The radial equation then becomes
\begin{equation}
\left[-\frac{1}{2M}\left(\partial_k^2 + \frac{\partial_k}{k_0}\right) + 2(\delta k)^2\right]\psi(\k) =E_0\,\psi(\k).
\end{equation}
At leading order, we neglect the first-derivative term in the radial Laplacian -- this approximation will be justified a posteriori. The radial equation then reduces to
\begin{equation}
\left[
-\frac{\partial_{k}^2}{2M}
+
2(\delta k)^2
\right]\psi(\k)
=
E_0\,\psi(\k).
\end{equation}
This is a one-dimensional harmonic oscillator in the radial momentum coordinate $\delta k$, with ground state 
\begin{equation}\label{eq:kspace_single_particle_HO}
\psi_0(k)
=
\left(\frac{2\sqrt{M}}{\pi}\right)^{1/4}
\exp\left[
-(k-k_0)^2/2\sigma_k^2
\right],
\end{equation}

with Gaussian width defined as
\begin{equation}
\sigma_k^2=\frac{1}{2\sqrt{M}}=\frac{k_0\omega}{2}.
\end{equation}
The ground state is therefore a Gaussian annulus in momentum space, localized around the circle $k=k_0$ and uniform along the angular direction.

We can now check the approximation made above when neglecting the linear derivative $\propto \partial_k /k_0$. The neglected term scales like
\begin{equation}
\frac{\partial_{k}}{k_0} \sim \frac{\sigma_k}{k_0} = \sqrt{\frac{\omega}{2k_0}}.
\end{equation}
The approximation is therefore controlled when $\omega\ll k_0$, i.e., when the inverse lengthscale $\sqrt{\Omega^2/\gamma k_0^4}$ is much smaller than the Mexican hat ring momentum $k_0$.
This is precisely the limit in which the wavefunction forms a thin annulus in momentum space around the Mexican-hat minimum.

Let's now transform the ground state wave function back to real space.
Since the momentum-space ground state is rotationally invariant, its Fourier transform only depends on $r=|\r|$. Up to an overall normalization,
\begin{equation}
\Psi_0(\r)
=
\int \frac{d^2 \k}{(2\pi)^2}
e^{i\k\cdot\r}
\Psi_0(\k)
=
\int_0^\infty k\,dk\,
\psi_0(k) J_0(kr),
\end{equation}
where we used
\begin{equation}
\int_0^{2\pi}d\phi\, e^{ikr\cos\phi}=2\pi J_0(kr).
\end{equation}
Substituting the Gaussian form of the momentum-space wavefunction gives
\begin{equation}
\Phi_0(r)
\propto
\int_0^\infty k\,dk\,
\exp\left[
-\frac{(k-k_0)^2}{2\sigma_k^2}
\right]
J_0(kr).
\end{equation}
In the thin-annulus limit $\sigma_k\ll k_0$, the integral is dominated by momenta close to $k_0$, and the Bessel function can be pulled out from the integral. The real-space wavefunction then takes the approximate form
\begin{equation}\label{eq:realspace_single_particle_HO}
\Phi_0(r)
\propto
e^{-\sigma_k^2 r^2/2}
J_0(k_0 r),
\end{equation}
up to slowly varying prefactors. Thus, unlike the usual harmonic oscillator with quadratic dispersion, the ground state is not a simple Gaussian in real space. Instead, it oscillates on the scale $1/k_0$, set by the radius of the Mexican-hat minimum, while its envelope varies on the longer scale
\begin{equation}
\xi \sim \frac{1}{\sigma_k}
=
\sqrt{\frac{2}{k_0\omega}}.
\end{equation}
Crucially, the real-space wave orbital will display nodes at the zeroes of the Bessel function, with the first zero located at $r \approx 2.4048/k_0$ -- this matches quantitatively with the radius of the nodal rings shown in Fig.\ \ref{fig:NNplots} in the main text.

\end{document}